# Entropy spectroscopy of a tunable two-site Hubbard molecule

Uhjin Kim[1*], Seokyeong Lee[2*], Gibum Yun[2*], Dongsung T. Park[3], Soobeom Choi[1], Sangwoo Jeong[1], Donghoon Kim[2#], V. Umansky[4], Yunchul Chung[5†], Hyoungsoon Choi[2,6‡], H.-S. Sim[2§], Hyung Kook Choi[1¶]

*[1] Department of Physics, Research Institute of Physics and Chemistry, Jeonbuk National University, Jeonju 54896, Republic of Korea.*

*[2] Department of Physics, KAIST, Daejeon 34141, Republic of Korea.*

*[3] Department of Physics, POSTECH, Pohang 37673, Republic of Korea.*

*[4] Department of Condensed Matter Physics, Weizmann Institute of Science, Rehovot, 76100, Israel.*

*[5] Department of Physics, Pusan National University, Busan 46241, Republic of Korea.*

*[6] Graduate School of Quantum Science and Technology, KAIST, Daejeon 34141, Republic of Korea.*

[*] These authors contributed equally: U.K., S.L., and G.Y.

[†] ycchung@pusan.ac.kr

[‡] h.choi@kaist.ac.kr

[§] hs_sim@kaist.ac.kr

[¶] hkchoi@jbnu.ac.kr

[#] Present Address: Analytical Quantum Complexity RIKEN Hakubi Research Team, RIKEN Center for Quantum Computing (RQC), Wako, Saitama 351-0198, Japan.

**Abstract**

Determining which microscopic states remain thermally active when tunneling and interactions compete is a central question in Hubbard physics. Here we develop an entropy measurement protocol for a double quantum dot, extending the charge-based approach established for a single dot, and measure the total entropy of a tunable two-site Hubbard molecule realized in a GaAs double dot. By shifting both dot levels together at fixed detuning, the entropy is correctly probed with only a single charge sensor. As interdot tunneling increases, the system evolves from two atomic-like dots through hybridized molecular states to a merged single dot. Across this evolution, the entropy decreases as tunnel-induced energy splitting exceeds the thermal energy and suppresses the occupation of higher-energy states. The measurements resolve the diminishing contribution of antibonding states and the changing thermal contributions of hybridized singlet and triplet states, in quantitative agreement with a two-site Hubbard model. By distinguishing states with the same charge configuration but different orbital and spin content, entropy reveals how tunneling and interactions determine the thermally active states of the minimal Hubbard system.

## Main

Entropy reflects the number of microscopic states in thermal equilibrium, providing direct insight into ground-state degeneracy and thermally accessible excited states[1–10]. Conventional calorimetry, however, remains impractical at the quantum-device scale because the heat involved is exceedingly small and rapidly dissipates into the environment[3,4,6,7]. An alternative approach exploits a Maxwell relation linking entropy changes to the temperature dependence of charge occupation, enabling direct observation of entropy arising from electron spin degeneracy[4,7,9]. The method, realized with a single quantum dot (SQD), has since been extended to SQDs strongly coupled to reservoirs[5], including those in the Kondo regime[11–13], and to valley and orbital degeneracies in bilayer graphene quantum dots (QDs)[14]. Together with recent advances in gate-defined QD arrays as tunable simulators of Fermi–Hubbard physics[15–20], these developments motivate extending entropy measurements to multidot systems[21–23]. As a first step in this direction, a double quantum dot (DQD) provides the minimal bridge between individual QDs and larger QD simulators and is also a controllable realization of a two-site Hubbard molecule[24–26]. On-site and inter-site Coulomb interactions, together with tunable interdot hopping, generate bonding and antibonding orbitals and spin singlet and triplet states[21,22,27–31]. Their evolving degeneracies should produce distinct entropy signatures, including a reduction relative to two independent dots, but these have not yet been resolved experimentally.

Extending charge-based entropy measurements from an SQD to a DQD introduces a fundamental measurement challenge: the thermodynamic requirement of the Maxwell relation and the experimental readout do not generally coincide. This mismatch originates from the two charge degrees of freedom in a DQD: the Maxwell relation requires a specific combination of both QD occupations, whereas a charge sensor (CS) provides a generally different combination determined by its coupling to the two QDs. A recent report[23] extended the approach to a DQD, but the inferred entropy exhibits clear discrepancies with the theoretically expected values in parts of the charge stability diagram, highlighting how this mismatch limits the direct extension of the SQD measurement scheme to larger QD systems. One possibility is to employ multiple charge sensors to measure the occupation of each QD separately[32]. This strategy, however, becomes impractical as the array grows owing to increasing device complexity and cross-capacitive coupling[16,19,33]. A general method should therefore access the relevant charge information with a minimal CS configuration, yet how to achieve this remains an open question.

Here we establish such a protocol for measuring the total entropy of a DQD by shifting both QD levels together at fixed detuning, thereby defining a $\delta$-axis. Along the $\delta$-axis, the Maxwell relation depends only on the total DQD occupation, which can be measured using a single CS coupled to both QDs without resolving their individual occupations. Numerical calculations show that applying the Maxwell relation along the $\delta$-axis reproduces the Gibbs entropy across the charge stability diagram, whereas applying it along an individual plunger-gate direction can yield apparent entropy that differs from the Gibbs entropy.

Experimentally, we use this protocol to measure the entropy of a tunable two-site Hubbard molecule as increasing interdot tunnel coupling drives it from two atomic-like dots through hybridized molecular states to a merged-dot regime. In the one-electron region, the entropy decreases with increasing interdot tunnel coupling as the bonding–antibonding splitting exceeds the thermal energy, suppressing occupation of the antibonding states. In the two-electron region, the entropy similarly captures the changing thermal contributions of the singlet and triplet states. In particular, stronger hybridization between the singlet states increases their energy splitting, suppressing the thermal contribution of the antibonding singlet state and thereby reducing the entropy. More broadly, our protocol establishes a thermodynamically consistent framework for entropy spectroscopy of coupled QD systems and provides a practical route toward larger QD-based Hubbard simulators.

**Device Characterization**

A false-colored scanning electron microscope (SEM) image of the device is shown in Fig. 1(a). The device consists of a serial DQD, two independent heating reservoirs, and a nearby charge sensor (CS, green) implemented as a quantum point contact (QPC). The energy level of each QD, $\varepsilon_{\mathrm{L}}$ or $\varepsilon_{\mathrm{R}}$, is controlled by its corresponding plunger gates, the left ($V_{\mathrm{LP}}$, purple) and right ($V_{\mathrm{RP}}$, purple), while cross-capacitance introduces a weak influence on the opposite QD, characterized by the lever-arm coefficients $\alpha_{ij} \equiv -\frac{\partial \varepsilon_i}{\partial V_j}$ [21,34], where $i, j \in \{\mathrm{L}, \mathrm{R}\}$ denote the gate indices (Supplementary Note 1). When the energy level of a QD aligns with the Fermi level of the corresponding reservoir, an electron tunnels into the QD, altering the local electrostatic potential. This change is capacitively coupled to the nearby CS, manifesting as a distinct kink in its conductance signal, $G \equiv I_{\mathrm{CS}}/V_{\mathrm{AC}}$, upon each change in the DQD charge occupancy[29,32,35]. The current path for the CS QPC is drawn with a blue solid arrow. The advantage of our device is that the DQD can be characterized via both direct transport measurements and charge sensing, noting that the data presented here are primarily obtained using the CS. Throughout the measurements, the reservoir coupling strength $\Gamma$ was kept much smaller than the thermal energy scale $k_{\mathrm{B}}T_{\mathrm{e}}$, to prevent the formation of unwanted hybridized states between the reservoir and the QDs[5,11,36] (Supplementary Note 2).

A key feature of the device is the wide tunability of the interdot tunnel coupling $t_{\mathrm{c}}$ via the middle-gate voltage $V_{\mathrm{MG}}$ (the red false-colored gate in the inset of Fig. 1(b)). The DQD can be continuously tuned from an atomic-like regime of two nearly independent QDs to a hybridized molecular regime. To quantify $t_{\mathrm{c}}$, we measured the CS conductance $G$ and fitted it to the mean population $n_{\mathrm{L}}(\varepsilon)$ [32,33] across the interdot charge transition $(N_{\mathrm{L}}, N_{\mathrm{R}}+1) \rightarrow (N_{\mathrm{L}}+1, N_{\mathrm{R}})$ while sweeping the gate voltages along the detuning axis, $\varepsilon \equiv \varepsilon_{\mathrm{R}} - \varepsilon_{\mathrm{L}}$ (Supplementary Note 3). Here, $N_{\mathrm{L}}$ and $N_{\mathrm{R}}$ are the occupation numbers of the left and right QDs of the DQD, respectively. For each $V_{\mathrm{MG}}$, the extracted $t_{\mathrm{c}}$ is plotted in Fig. 1(b), demonstrating that $t_{\mathrm{c}}$ can be tuned over a wide range from below $20$ μeV to beyond $300$ μeV.

Figure 1(c) shows the electron temperatures $T_{\mathrm{e}}$ of the left QD (L-QD, orange) and right QD (R-QD, purple) as a function of heating current $I_{\mathrm{heat}}$. For each QD, the reservoirs consist of two closed chambers, which we define as the direct and indirect heating chambers[8]. A heating current $I_{\mathrm{heat}}$ (white dashed arrow in Fig. 1(a)) flows across the direct heating chamber through two QPCs each connecting to the source and drain of the current, respectively. This way, the direct chamber heats up without an unwanted bias, and only the heat is transferred to the indirect heating chamber (red arrow). $T_{\mathrm{e}}$ was extracted from the thermal broadening $\theta$ of the CS signal across a charge transition, where $\theta \equiv \frac{k_{\mathrm{B}}T_{\mathrm{e}}}{\alpha}$ is in units of voltage[4,32]. Without heating, $T_{\mathrm{e}}$ is approximately $200$ mK, consistent with the value independently obtained from Coulomb blockade thermometry using an SQD (Supplementary Note 2). As $I_{\mathrm{heat}}$ increases, both $T_{\mathrm{e,L}}$ and $T_{\mathrm{e,R}}$ rise monotonically, reaching nearly $1$ K at the maximum applied current ($I_{\mathrm{heat}} \approx 20$ nA). The minute temperature difference between the two reservoirs, which is about $10$ mK, confirms that no significant thermal imbalance was generated. At low heating currents, the temperature follows the expected quadratic dependence $T_{\mathrm{e}} \propto I_{\mathrm{heat}}^2$ (black dashed curve), consistent with Joule heating power $P \propto I^2$ [8,37,38]. Above $10$ nA, $T_{\mathrm{e}}$ deviates from the quadratic trend, which can be attributed to enhanced electron–phonon coupling[8,39,40]. The inset of Fig. 1(c) shows the thermal broadening $\theta$ for the R-QD charge transition with increasing $I_{\mathrm{heat}}$. The peak position remains unchanged, confirming that the heating raises $T_{\mathrm{e}}$ without shifting the reservoir chemical potentials (Supplementary Note 4).

**DQD Entropy Measurement Protocol**

The entropy $S$ of an SQD has been extracted via the Maxwell relation $\frac{dS}{d\mu} = \frac{dN}{dT}$ [4,5,7,8]. One must take caution in extending this approach to a DQD with a single CS[23]. The complication stems from the two charge degrees of freedom, $N_{\mathrm{L}}$ and $N_{\mathrm{R}}$, which are cross-controlled by two plunger gate voltages such as $V_{\mathrm{LP}}$ and $V_{\mathrm{RP}}$ in our setup[21,34]. Ideally, $V_{\mathrm{L(R)P}}$ would only control $N_{\mathrm{L(R)}}$, but in reality, tuning of a single plunger can alter the charge states of both QDs. As a result, the charge information required by the Maxwell relation differs and may not be accessible with a single CS. In the case that only one plunger gate voltage, $V_{\mathrm{LP}}$, is swept for the entropy measurement, as is done for an SQD, the corresponding Maxwell relation is

$$\left.\frac{\partial S}{\partial V_{\mathrm{LP}}}\right|_{T_{\mathrm{e}},V_{\mathrm{RP}}} = \left.\frac{\partial(\alpha_{\mathrm{LL}}N_{\mathrm{L}} + \alpha_{\mathrm{RL}}N_{\mathrm{R}})}{\partial T_{\mathrm{e}}}\right|_{V_{\mathrm{LP}},V_{\mathrm{RP}}} \quad (1)$$

(Supplementary Note 5). The charge information required for this relation is $\alpha_{\mathrm{LL}}N_{\mathrm{L}} + \alpha_{\mathrm{RL}}N_{\mathrm{R}}$. Because changes in $N_{\mathrm{L}}$ and $N_{\mathrm{R}}$ are combined in the CS signal[32], a single CS cannot resolve the two occupations separately when they vary simultaneously, particularly in the vicinity of the triple points. Resolving the individual occupations would require two distinct CSs, introducing additional experimental complexities associated with calibrating the intricate cross-capacitance matrix among the two QDs and the two sensors[16,33].

We address this problem by simultaneously varying $V_{\mathrm{LP}}$ and $V_{\mathrm{RP}}$ along a specific trajectory that shifts both QD energy levels by the same amount $d\delta$, thereby keeping the energy difference $\varepsilon = \varepsilon_{\mathrm{R}} - \varepsilon_{\mathrm{L}}$ constant. We refer to this as the "$\delta$-axis" trajectory, which runs parallel to the charge degeneracy line between $(1,0)$ and $(0,1)$ in the stability diagram[21].

Along the $\delta$-axis trajectory, the Maxwell relation can be written in terms of the total occupation $N_{\mathrm{DQD}} = N_{\mathrm{L}} + N_{\mathrm{R}}$ of the DQD:

$$\left.\frac{\partial S}{\partial \delta}\right|_{T_{\mathrm{e}},\varepsilon} = -\left.\frac{\partial N_{\mathrm{DQD}}}{\partial T_{\mathrm{e}}}\right|_{\delta,\varepsilon} \tag{2}$$

(Supplementary Note 5). Importantly, equation (2) is independent of $\alpha_{ij}$. Along the $\delta$-axis, the fixed detuning keeps the relative charge distribution between the two QDs unchanged for a given $N_{\mathrm{DQD}}$ and $T_{\mathrm{e}}$. The CS signal therefore directly tracks changes in the total occupation, allowing $N_{\mathrm{DQD}}$ to be measured with a single CS without separately determining $N_{\mathrm{L}}$ and $N_{\mathrm{R}}$. We first validated the $\delta$-axis protocol numerically using a DQD Hamiltonian. We calculated $N_{\mathrm{L}}$ and $N_{\mathrm{R}}$ as functions of $V_{\mathrm{LP}}$ and $V_{\mathrm{RP}}$ at finite temperature and used them to simulate the CS conductance $G$. Entropy was then extracted from the simulated $G$ using equation (2) along the $\delta$-axis trajectories and, for comparison, equation (1) along the $V_{\mathrm{LP}}$-axis trajectories. Details of the models and numerical procedure are provided in Methods.

Figure 2(a) presents the schematic charge stability diagram used in the simulations, including the charge configurations $(N_{\mathrm{L}}, N_{\mathrm{R}})$, the electron triple point (ETP, orange circle), and the hole triple point (HTP, blue circle). The corresponding Gibbs entropy is plotted in Fig. 2(b). The entropy extracted along $\delta$-axis trajectories (Fig. 2(c)) reproduces the Gibbs entropy throughout the charge stability diagram. In contrast, the entropy extracted along $V_{\mathrm{LP}}$-axis trajectories (Fig. 2(d)) exhibits clear deviations from the Gibbs entropy, with spurious lines extending from the ETP and HTP parallel to the $V_{\mathrm{LP}}$-axis. This comparison demonstrates that the $\delta$-axis protocol enables reliable extraction of the total DQD entropy using a single CS.

**Entropy Evolution Across the $(0,0) \rightarrow (1,0) \rightarrow (1,1)$ Regions**

We apply the $\delta$-axis protocol in the regime away from the ETP and HTP. Figure 3(a) shows the charge stability diagram of the DQD, with the $\varepsilon$- and $\delta$-axes and the entropy measurement path indicated by the black dashed arrow. Along this path, $\varepsilon \sim 300$ μeV well exceeds the average thermal energy scale $k_{\mathrm{B}}\bar{T}_{\mathrm{e}} \sim 40$ μeV, where $\bar{T}_{\mathrm{e}} \equiv \left(T_{\mathrm{e,c}} + T_{\mathrm{e,h}}\right)/2$ is the average temperature of the unheated $(I_{\mathrm{heat}} = 0\ \mathrm{nA}, T_{\mathrm{e,c}})$ and heated (10 nA, $T_{e,h}$) cases, and interdot tunnel coupling $t_c$~20 μeV, suppressing interdot hybridization so that the two QDs behave effectively as independent SQDs[21,23].

Figure 3(b) shows the CS conductance $G$ measured along the path indicated in Fig. 3(a), corresponding to the $(0,0) \rightarrow (1,0) \rightarrow (1,1)$ transition sequence. The blue circles and red squares represent the raw conductance data for the unheated ($I_{\mathrm{heat}} = 0$ nA) and heated (10 nA) cases, respectively, while the solid lines denote the averaged traces. Fitting the conductance traces gives the thermal broadening $\theta$. Together with the previously measured lever-arm values, this yields the temperature rise $\Delta T_e \approx 300$ mK. The averaged signals are then normalized to obtain $N_{\mathrm{DQD}}$ shown in Fig. 3(c), from which the entropy along the trace is calculated using equation (2) (see Supplementary Note 6). The entropy evolution in Fig. 3(d) follows the spin and charge degeneracies expected for the two independent QDs, consistent with prior reports[4,5,23]. Starting from the empty state (0,0) with $\Delta S = k_B \ln 1$, the system reaches $\Delta S = k_B \ln 3$ at the midpoint of the $(0,0) \rightarrow (1,0)$ transition, where the three states $\{|0,0\rangle, |\uparrow,0\rangle, |\downarrow,0\rangle\}$ become equally accessible. After the transition, only the L-QD spin degeneracy remains with $\Delta S = k_B \ln 2$. At the midpoint of the $(1,0) \rightarrow (1,1)$ transition, the two spin states of the (1,0) and the four spin states of the (1,1) become equally accessible, $\{|\uparrow,0\rangle, |\downarrow,0\rangle, |\uparrow,\downarrow\rangle, |\uparrow,\uparrow\rangle, |\downarrow,\uparrow\rangle, |\downarrow,\downarrow\rangle\}$, resulting in $\Delta S = k_B \ln 6$. Finally, the entropy saturates at $\Delta S = k_B \ln 4$ in the (1,1) region. These observations are consistent with Fig. 2(c).

**Entropy Evolution of a Hubbard Molecule with Interdot Tunnel Coupling $t_c$**

For reasons discussed above, we measure the entropy along the δ-axis through both the ETP and HTP for different strengths of the interdot tunnel coupling $t_c$, following the system from two nearly independent QDs at weak tunneling, through a hybridized molecular regime, to a merged-SQD limit at strong tunneling (Fig. 4).

In the weak-tunneling regime with $t_c < k_B \bar{T}_e$ (Fig. 4(a–c)), the DQD retains an atomic-like character with negligible interdot hybridization. Figure 4(a) presents the measured charge stability diagram. Although the actual charge configurations are $(N_L, N_R) = (4,10), (4,11), (5,10)$, and $(5,11)$, we use the effective charge labels as $(0,0), (0,1), (1,0)$, and $(1,1)$, as depicted in Fig. 4(a). The asymmetric confinement potential of the DQD lifts additional orbital degeneracies, and bias spectroscopy confirms that the observed excited state lies well above the thermal energy scale (Supplementary Note 1), leaving only charge and spin as the relevant degrees of freedom[22,41]. A fit to the mean L-QD population $n_L(\varepsilon)$ along the $\varepsilon$-axis gives $t_c$~ 6.5 μeV (Fig. 4(b)). Although $t_c$ is not precisely resolved in this regime, the result is consistent with $t_c < k_B \bar{T}_e \sim 40$ μeV. Figure 4(c) displays the measured entropy (blue circles) together with the Gibbs entropy (blue solid line).

In the (0,0) region, the measured entropy is $k_{\mathrm{B}} \ln 1$ as expected for the empty state $\{|0,0\rangle\}$. At the boundary between the (0,1) and (1,0) regions, interdot tunneling mixes the two charge states to form two low-energy bonding states $\{|\uparrow,0\rangle - |0,\uparrow\rangle, |\downarrow,0\rangle - |0,\downarrow\rangle\}$, and two high-energy antibonding states $\{|\uparrow,0\rangle + |0,\uparrow\rangle, |\downarrow,0\rangle + |0,\downarrow\rangle\}$, with an energy splitting of $2t_{\mathrm{c}}$. Because $2t_{\mathrm{c}} < k_{\mathrm{B}}\bar{T}_{\mathrm{e}}$, all four states are thermally populated, yielding $\Delta S = k_{\mathrm{B}} \ln 4$. Deep in the (1,1) region, spin-exchange interactions separate the two-electron states into a low-energy singlet state $\{|\uparrow,\downarrow\rangle - |\downarrow,\uparrow\rangle\}$ and a set of high-energy triplet states $\{|\uparrow,\downarrow\rangle + |\downarrow,\uparrow\rangle, |\uparrow,\uparrow\rangle, |\downarrow,\downarrow\rangle\}$, with exchange energy[24,31] $J = \frac{4t_{\mathrm{c}}^2}{(E_{\mathrm{C}} - E_{\mathrm{Cm}})}$, where $E_{\mathrm{C}}$ is the charging energy of each QD and $E_{\mathrm{Cm}}$ is the mutual charging energy[22,24,25,28]. Because $J \sim 0.05$ μeV is much smaller than the thermal energy $k_{\mathrm{B}}\bar{T}_{\mathrm{e}} \sim 40$ μeV, the singlet and triplet states are all thermally populated, again giving $\Delta S = k_{\mathrm{B}} \ln 4$. At the ETP (orange circle) and HTP (green circle), the entropy reflects the combined number of thermally populated microstates from the neighboring regions, giving an entropy of $k_{\mathrm{B}} \ln 5$ and $k_{\mathrm{B}} \ln 8$, respectively. Thus, in the weak-tunneling regime, both energy splittings ($2t_{\mathrm{c}}$ and $J$) remain below $k_{\mathrm{B}}\bar{T}_{\mathrm{e}}$, keeping all of the bonding, antibonding, singlet and triplet states thermally accessible.

At intermediate tunneling ($t_{\mathrm{c}} > k_{\mathrm{B}}\bar{T}_{\mathrm{e}}$) (Fig. 4(d–f)), interdot hybridization becomes substantial[21]. Figures 4(d) and 4(e) present the corresponding charge stability diagram and $t_{\mathrm{c}}$ measurement, respectively. Here, $t_{\mathrm{c}}(\sim 210$ μeV) is more than five times $k_{\mathrm{B}}\bar{T}_{\mathrm{e}}$. The entropy evolution in Fig. 4(f) (experiment: orange circles; theory: solid line) exhibits a clear reduction relative to the weak-tunneling regime in Fig. 4(c). At the boundary between the (0,1) and (1,0) regions, the splitting $2t_{\mathrm{c}}$ exceeds $k_{\mathrm{B}}\bar{T}_{\mathrm{e}}$, suppressing occupation of the antibonding states and leaving only two bonding states thermally accessible, yielding $\Delta S = k_{\mathrm{B}} \ln 2$. In the (1,1) region, the singlet–triplet exchange splitting $J \sim 58$ μeV is also larger than $k_{\mathrm{B}}\bar{T}_{\mathrm{e}}$ [28], reducing the contribution of the triplet states to the entropy, leading to $\Delta S \sim k_{\mathrm{B}} \ln 3$, somewhat reduced from $k_{\mathrm{B}} \ln 4$. At the ETP (cyan circle) and HTP (purple circle), the entropy again reflects the combined neighboring microstates. When $t_{\mathrm{c}}$ is increased further and becomes comparable to the charging energy, the system approaches the merged-SQD limit (Fig. 4(g–i)). Accordingly, the CS conductance no longer distinguishes the two charge states (Fig. 4(h)). In this limit, the measured entropy (Fig. 4(i), green circles) follows the behavior expected for sequential one- and two-electron filling of an SQD, consistent with the SQD model[21,22] calculation (green solid line, Supplementary Note 7) and prior measurements[4].

**Entropy Signatures of One- and Two-Electron Hubbard Molecular States**

Having established the entropy measurement across the full range of $t_{\mathrm{c}}$, we now examine how the internal eigenstates of the DQD evolve from localized atomic states to hybridized molecular states within fixed $N_{\mathrm{DQD}}$ regions. We trace the $\varepsilon$-dependent entropy across the $(1,0)$–$(0,1)$ one-electron region and the $(0,2)$–$(1,1)$ two-electron region for different $t_{\mathrm{c}}$, using a series of $\delta$-axis line cuts.

Figure 5(a) presents the absolute entropy across the $(0,1) \rightarrow (1,0)$ transition $S(\varepsilon) \equiv \Delta S(\varepsilon) + S(0,0) = \Delta S(\varepsilon)$ for two values of $t_{\mathrm{c}}$ in the $N_{\mathrm{DQD}} = 1$ region. For these measurements, the heating current was increased from 10 to 20 nA to improve the signal-to-noise ratio; this larger temperature modulation produced no discernible change in the extracted entropy (Supplementary Note 8). The red filled circles and red dashed line represent the experimental data and theoretical calculation for $t_{\mathrm{c}} = 6.9$ µeV, whereas the blue open circles (experiment) and blue dotted line (theory) correspond to $t_{\mathrm{c}} = 86.4$ µeV. At each $\varepsilon$, $\Delta S(\varepsilon)$ is determined from the local minimum value in the $N_{\mathrm{DQD}} = 1$ region as indicated by the red dot in the inset of Fig. 5(a). The full two-dimensional entropy evolution maps are provided in Supplementary Note 9. The full widths at half maximum (FWHMs) of both theoretical curves are shaded in their corresponding colors and marked by vertical lines. Figure 5(b) shows the corresponding schematic energy-level diagram for the $N_{\mathrm{DQD}} = 1$ region. At large $|\varepsilon|$ (left and right sides of Fig. 5(b)), the entropy approaches $k_{\mathrm{B}} \ln 2$, consistent with a single electron localized in either the $|1,0\rangle$ or $|0,1\rangle$ state with spin degeneracy[4]. As $|\varepsilon|$ decreases towards zero (center of Fig. 5(b)), the two localized states approach in energy and hybridize into bonding and antibonding states. Thus, the magnitude of the entropy feature reflects the number of thermally accessible states, while its width along the $\varepsilon$-axis qualitatively characterizes the detuning range over which the two QD levels are sufficiently close to hybridize (Supplementary Note 10). In the weak-tunneling regime ($t_{\mathrm{c}} = 6.9$ µeV), the entropy develops a peak approaching $\Delta S(\varepsilon) = k_{\mathrm{B}} \ln 4$ near $\varepsilon = 0$ [27]. Increasing $t_{\mathrm{c}}$ to 86.4 µeV broadens the entropy feature along the $\varepsilon$-axis while reducing its magnitude near $\varepsilon = 0$. The broader feature reflects the increased detuning range of interdot hybridization, whereas the reduced magnitude results from the larger bonding–antibonding splitting and the diminished contribution of the antibonding states[32].

In the $(0,2)$–$(1,1)$ region, the entropy tracks the hybridization of the singlet states relative to the triplet states[24,28], for $t_{\mathrm{c}} = 14.4$ µeV (red filled squares) and $t_{\mathrm{c}} = 33.7$ µeV (blue open squares), together with the corresponding theoretical calculations (red dashed and blue dotted lines) as shown in Fig. 5(c). Because the initial $(0,1)$ state already carries a spin entropy of $k_B \ln 2$, we again plot the absolute entropy $S(\varepsilon) = \Delta S(\varepsilon) + S(0,1) = \Delta S(\varepsilon) + k_{\mathrm{B}} \ln 2$ . The corresponding schematic energy-level diagram is illustrated in Fig. 5(d). Deep in the $(0,2)$ region ($\varepsilon \ll 0$, left side of Fig. 5(d)), both the singlet and the triplet states are localized in the right QD. The typical energy splitting between the singlet state ($|S(0,2)\rangle$) and the triplet states ($\{T(0,2)\} \equiv \{|T_{+}(0,2)\rangle, |T_{0}(0,2)\rangle, |T_{-}(0,2)\rangle\}$), $\Delta E_{\mathrm{ST}}$, is on the order of a few hundred µeV [4,22,28–30]. Consistently, no excited states were observed within the applied bias range of our bias spectroscopy measurements (see Supplementary Note 1), leaving only the spin singlet state $|S(0,2)\rangle$ to be thermally populated, resulting in $S = k_{\mathrm{B}} \ln 1$. In contrast, in the $(1,1)$ region ($\varepsilon > 0$, right side of Fig. 5(d)) the energy splitting between the one singlet state ( $|S(1,1)\rangle$ ) and the three triplet states ( $\{T(1,1)\} \equiv \{|T_{+}(1,1)\rangle, |T_{0}(1,1)\rangle, |T_{-}(1,1)\rangle\}$) is sufficiently small ($J \ll k_{\mathrm{B}}\bar{T}_{\mathrm{e}}$) that all four states are nearly degenerate for both $t_{\mathrm{c}}$ values, leading to $S = k_{\mathrm{B}} \ln 4$. As $\varepsilon$ approaches zero (center of Fig. 5(d)), $|S(1,1)\rangle$ and $|S(0,2)\rangle$ hybridize to form bonding and antibonding singlets with a minimum splitting of $2\sqrt{2}t_{\mathrm{c}}$ [22,33,42] near the boundary between the $(0,2)$ and $(1,1)$ regions. The hybridization between the triplet and singlet states, $|S(0,2)\rangle \leftrightarrow \{T(1,1)\}$, is forbidden by spin selection rules[29,30,43], while $\{T(0,2)\}$ triplet states lie too high in energy to participate. Accordingly, the $\{T(1,1)\}$ triplet states remain in between the bonding and antibonding singlet states as shown in Fig. 5(d). When $t_{\mathrm{c}} = 14.4$ μeV, the minimum bonding–antibonding singlet splitting $2\sqrt{2}t_{\mathrm{c}}$ is comparable to $k_{\mathrm{B}}\bar{T}_{\mathrm{e}}$, allowing both hybridized singlet states and the three triplet states $\{T(1,1)\}$ to be populated, resulting in an entropy close to $S \sim k_{\mathrm{B}} \ln 5$. Increasing $t_{\mathrm{c}}$ to 33.7 μeV increases the splitting beyond $k_{\mathrm{B}}\bar{T}_{\mathrm{e}}$, reducing the contribution of the antibonding singlet and thereby suppressing the entropy enhancement to $S \sim k_{\mathrm{B}} \ln 4$.

**Discussion**

We have demonstrated entropy spectroscopy of a DQD across a broad range of charge configurations and interdot tunnel couplings $t_{\mathrm{c}}$. The measured entropy shows quantitative agreement with the Gibbs entropy calculated from a two-site Hubbard model. As $t_{\mathrm{c}}$ increases, the one-electron entropy captures both the broadened detuning range of interdot hybridization and the diminishing contribution of antibonding states, whereas the two-electron entropy tracks the hybridization of singlet states relative to the triplet states.

A key element of our entropy measurement is the identification of the $\delta$-axis, defined by uniformly shifting the energy levels of both QDs at fixed detuning. Along this direction, the Maxwell relation depends only on the total DQD occupation $N_{\mathrm{DQD}}$, allowing the total entropy to be extracted with a single CS without separately resolving $N_{\mathrm{L}}$ and $N_{\mathrm{R}}$. By linking the thermodynamic response directly to the total occupation, the $\delta$-axis provides a practical route for extending charge-based entropy measurements from a single QD to coupled QD systems[15–19,33].

This scalability opens new opportunities for entropy measurements in larger QD arrays operated as Fermi–Hubbard simulators[15–18]. Recent advances in such simulators have enabled the exploration of correlated phenomena, including finite-size Mott physics in a triple QD array[15,17] and Nagaoka ferromagnetism in a quadruple QD plaquette[18]. These systems motivate complementary thermodynamic probes sensitive to the thermal occupation of correlated many-body states. Tracking the evolution of entropy could provide thermodynamic signatures of local moment formation, degeneracy lifting and the associated quenching of low-energy degrees of freedom, thus offering an alternative perspective on correlated many-body states in mesoscopic quantum systems[2,4,5,11,14].

## Methods

### Experiment

The device was fabricated from a GaAs/AlGaAs heterostructure hosting a two-dimensional electron gas (2DEG) located approximately 75 nm below the surface. The 2DEG exhibits an electron density of $2.0 \times 10^{11}\ \mathrm{cm}^{-2}$ and a mobility of roughly $3.0 \times 10^{6}\ \mathrm{cm}^2\ \mathrm{V}^{-1}\ \mathrm{s}^{-1}$. The metallic surface gates defining the DQD were patterned by electron-beam lithography and metallized with Ti/Au using a standard lift-off process. Ohmic contacts to the 2DEG were fabricated by photolithography, metal deposition, and rapid thermal annealing to ensure low-resistance electrical access. All measurements were performed in a commercial dilution refrigerator with a base temperature of 50 mK. The effective electron temperature in the 2DEG, estimated from Coulomb blockade thermometry, was approximately 200 mK under base conditions.

An AC excitation of $V_{\mathrm{AC}} = 30\ \mu\mathrm{V}$ at 101 Hz was applied across the CS channel, and the resulting differential current $I_{\mathrm{CS}}$ was measured using a homemade current–voltage converter with a gain of $10^7\ \mathrm{V\ A}^{-1}$ [44]. The amplified signal was demodulated with a lock-in amplifier to extract the CS response, $G = I_{\mathrm{CS}}/V_{\mathrm{AC}}$.

### Theoretical Calculation

We describe the numerical simulations for the entropy maps shown in Fig. 2(c) and 2(d). We first computed the occupancies $N_{\mathrm{L}}$ and $N_{\mathrm{R}}$ and used them to simulate the charge sensor conductance $G$ based on the model relation[23] $G = -g_{\mathrm{R}}N_{\mathrm{R}} - g_{\mathrm{L}}N_{\mathrm{L}} + \gamma_{\mathrm{R}}V_{\mathrm{RP}} + \gamma_{\mathrm{L}}V_{\mathrm{LP}} + G_0$ between the conductance and the DQD occupancies. Here, $\gamma_{\mathrm{R(L)}}$ captures the capacitive response of the sensor to the right (left) plunger gate, and $G_0$ is a constant offset. Then, following the usual entropy measurement protocol[5], we extracted the temperature dependence of the charge information from the simulated conductance $G$ by fitting it to a thermally broadened charge transition conductance model $G_{\mathrm{fit}} \sim \tanh\left(\frac{V_{\mathrm{P}} - V_{\mathrm{c}}}{2\theta}\right)$. Although $N_{\mathrm{L}}$ and $N_{\mathrm{R}}$ had already been obtained, the charge information was extracted solely from the simulated conductance $G$ (mimicking experimental situations). Here $V_{\mathrm{P}}$ is a plunger gate voltage varied along a $\delta$-axis trajectory for the map in Fig. 2(c) or a $V_{\mathrm{LP}}$-axis trajectory for the map in Fig. 2(d), while $V_{\mathrm{c}}(\theta)$ is the center position of the charge transition and $\theta \propto k_{\mathrm{B}}T_{\mathrm{e}}$ represents the thermal broadening. The extracted charge information was combined with the Maxwell relations in equations (2) and (1) to obtain the entropy maps in Fig. 2(c) and 2(d), respectively. The details of the simulations are given below.

We numerically computed the DQD occupancies $N_{\mathrm{L}}$ and $N_{\mathrm{R}}$ as a function of the gate voltages $V_{\mathrm{LP}}$ and $V_{\mathrm{RP}}$ and the temperature $T_{\mathrm{e}}$ by employing the Hamiltonian[21] of the DQD, $\hat{H}_{\mathrm{DQD}} = \hat{H}_{\mathrm{R}} + \hat{H}_{\mathrm{L}} + \hat{H}_{\mathrm{C}} + \hat{H}_{t_{\mathrm{c}}}$.

Here $\hat{H}_{\mathrm{L(R)}} = \sum_{\sigma=\uparrow,\downarrow} \varepsilon_{\mathrm{L(R)}} \hat{c}^{\dagger}_{\mathrm{L(R)},\sigma} \hat{c}_{\mathrm{L(R)},\sigma} + E_{\mathrm{C,L(R)}}\, \hat{n}_{\mathrm{L(R)},\uparrow} \hat{n}_{\mathrm{L(R)},\downarrow}$ describes the L-QD (R-QD). Each QD was modeled as having a single level $\varepsilon_{\mathrm{L(R)}}$, as we focused on the charge regions of $N_{\mathrm{DQD}} = 0,1,2$. The effective energy levels, $\varepsilon_{\mathrm{L}} = \frac{E_{\mathrm{C,L}}}{2} - \alpha_{\mathrm{LL}}(V_{\mathrm{LP}} - V_{\mathrm{L0}}) - \alpha_{\mathrm{LR}}(V_{\mathrm{RP}} - V_{\mathrm{R0}})$ and $\varepsilon_{\mathrm{R}} = \frac{E_{\mathrm{C,R}}}{2} - \alpha_{\mathrm{RR}}(V_{\mathrm{RP}} - V_{\mathrm{R0}}) - \alpha_{\mathrm{RL}}(V_{\mathrm{LP}} - V_{\mathrm{L0}})$, are controlled by the plunger gates $V_{\mathrm{LP}}$ and $V_{\mathrm{RP}}$, while $E_{\mathrm{C,L(R)}}$ is the charging energy of the L-QD (R-QD). The operator $\hat{c}^{\dagger}_{\mathrm{L(R)},\sigma}$ creates an electron with spin $\sigma$ in the energy level $\varepsilon_{\mathrm{L(R)}}$, and $\hat{n}_{\mathrm{L(R)},\sigma} = \hat{c}^{\dagger}_{\mathrm{L(R)},\sigma} \hat{c}_{\mathrm{L(R)},\sigma}$ is the number operator. The gate voltage offsets, $V_{\mathrm{L0}}$ and $V_{\mathrm{R0}}$, set $\varepsilon_{\mathrm{L}}$ and $\varepsilon_{\mathrm{R}}$ relative to the reservoir chemical potential, which is taken as zero. $\hat{H}_{\mathrm{C}} = E_{\mathrm{Cm}} \hat{n}_{\mathrm{L}} \hat{n}_{\mathrm{R}}$ describes the interdot Coulomb repulsion with the interaction strength $E_{\mathrm{Cm}}$, where $\hat{n}_{\mathrm{L(R)}} = \sum_{\sigma=\uparrow,\downarrow} \hat{n}_{\mathrm{L(R)},\sigma}$. The interdot tunneling term is written as $\hat{H}_{t_{\mathrm{c}}} = t_{\mathrm{c}} \sum_{\sigma=\uparrow,\downarrow} \left( \hat{c}^{\dagger}_{\mathrm{L},\sigma} \hat{c}_{\mathrm{R},\sigma} + \mathrm{h.c.} \right)$, where $t_{\mathrm{c}}$ is the interdot tunnel coupling. The DQD was assumed to remain in thermal equilibrium with the reservoirs, while hybridization of each QD with its reservoir was neglected because the QD–reservoir couplings are weak. The average charge occupancies $N_{\mathrm{R(L)}}(V_{\mathrm{LP}}, V_{\mathrm{RP}}; T_{\mathrm{e}}) = \mathrm{Tr}_{\mathrm{DQD}}\left[ \hat{n}_{\mathrm{R(L)}} \hat{\rho}_{\mathrm{DQD}} \right]$ were calculated, where $\hat{\rho}_{\mathrm{DQD}}\, (V_{\mathrm{LP}}, V_{\mathrm{RP}}; T_{\mathrm{e}}) = \frac{e^{-\beta \hat{H}_{\mathrm{DQD}}\, (V_{\mathrm{LP}}, V_{\mathrm{RP}})}}{Z(V_{\mathrm{LP}}, V_{\mathrm{RP}}; T_{\mathrm{e}})}$ is the DQD density matrix, $Z(V_{\mathrm{LP}}, V_{\mathrm{RP}}; T_{\mathrm{e}}) = \mathrm{Tr}_{\mathrm{DQD}}\left[ e^{-\beta \hat{H}_{\mathrm{DQD}}} \right]$ is the partition function, $\beta = 1/(k_{\mathrm{B}} T_{\mathrm{e}})$, and $\mathrm{Tr}_{\mathrm{DQD}}$ denotes the trace over all states of the DQD.

From the computed $N_{\mathrm{L}}$ and $N_{\mathrm{R}}$, we obtained the simulated conductance $G$ as a function of the gate voltages and the temperature. As in entropy measurements[5], the simulated conductance $G$ was fitted to a charge transition model[4], where the charge number and the model conductance follow $N = \frac{1}{2} \tanh\left( \frac{V_{\mathrm{P}} - V_{\mathrm{c}}}{2\theta} \right) + \frac{1}{2}$ and $G_{\mathrm{fit}} = -\frac{I}{2} \tanh\left( \frac{V_{\mathrm{P}} - V_{\mathrm{c}}}{2\theta} \right) + \gamma_1 V_{\mathrm{P}} + \gamma_2$, while the charge number changes by one as a gate voltage $V_{\mathrm{P}}$ varies across $V_{\mathrm{c}}$. Here, $I$ quantifies the conductance drop at the charge transition, $\gamma_1$ captures the cross-capacitance between the CS and the gate, and $\gamma_2$ is an offset. The thermal broadening $\theta \propto k_{\mathrm{B}} T_{\mathrm{e}}$ of the transition is determined by the lever-arm coefficients, depending on a selected trajectory in the gate-voltage space along which the entropy is measured. For a $\delta$-axis trajectory (along which both $V_{\mathrm{LP}}$ and $V_{\mathrm{RP}}$ are tuned to follow $V_{\mathrm{RP}} = \frac{\alpha_{\mathrm{LL}} - \alpha_{\mathrm{RL}}}{\alpha_{\mathrm{RR}} - \alpha_{\mathrm{LR}}} V_{\mathrm{LP}} - \frac{\varepsilon}{\alpha_{\mathrm{RR}} - \alpha_{\mathrm{LR}}}$), $\theta = k_{\mathrm{B}} T_{\mathrm{e}} \left( \frac{\alpha_{\mathrm{RR}} - \alpha_{\mathrm{LR}}}{\alpha_{\mathrm{RR}} \alpha_{\mathrm{LL}} - \alpha_{\mathrm{RL}} \alpha_{\mathrm{LR}}} \right)$ when $V_{\mathrm{P}} = V_{\mathrm{LP}}$ is chosen, since $d\delta = -\frac{\alpha_{\mathrm{RR}} \alpha_{\mathrm{LL}} - \alpha_{RL} \alpha_{\mathrm{LR}}}{\alpha_{\mathrm{RR}} - \alpha_{\mathrm{LR}}} dV_{\mathrm{LP}}$ (Supplementary Note 5). In this case, $N$ is identified as $N_{\mathrm{DQD}}$, according to the discussion around equation (2). On the other hand, for a $V_{\mathrm{LP}}$-axis trajectory (along which only $V_{\mathrm{LP}}$ is tuned), $V_{\mathrm{P}} = V_{\mathrm{LP}}$ must be chosen, and $\theta$ is found as $\theta = \frac{k_{\mathrm{B}} T_{\mathrm{e}}}{\alpha_{\mathrm{LL}}}$ when $N$ is identified as $N_{\mathrm{L}}$, while $\theta = \frac{k_{\mathrm{B}} T_{\mathrm{e}}}{\alpha_{\mathrm{RL}}}$ for $N = N_{\mathrm{R}}$ (see the discussion around equation (1)). By comparing the simulated conductance $G$ with the charge transition conductance model $G_{\mathrm{fit}}$, we extracted $\theta$ and $N$. Repeating this procedure at two adjacent temperatures $T_{\mathrm{e,c}}$ and $T_{\mathrm{e,h}}$, we evaluated $\frac{\partial N}{\partial \theta}$ at the average temperature $\bar{T}_{\mathrm{e}} = (T_{\mathrm{e,c}} + T_{\mathrm{e,h}})/2$. This approach was applied to the variation of the gate voltage $V_{\mathrm{P}}$ over a wide window where two successive charge transitions, $N=0\rightarrow 1$ and $N=1\rightarrow 2$, take place, utilizing an extended form of the transition conductance model, $G_{\mathrm{fit}}=-\frac{I_1}{2}\tanh\frac{V_{\mathrm{P}}-V_{\mathrm{c1}}}{2\theta_1}-\frac{I_2}{2}\tanh\frac{V_{\mathrm{P}}-V_{\mathrm{c2}}}{2\theta_2}+\gamma_1 V_{\mathrm{P}}+\gamma_2$.

The evaluated derivative $\frac{\partial N}{\partial\theta}$ was combined with a Maxwell relation to obtain the entropy. The Maxwell relation in equation (2) for a $\delta$-axis trajectory is rewritten as

$$\left.\frac{\partial S}{\partial V_{\mathrm{LP}}}\right|_{T_e}=\left(\frac{\alpha_{\mathrm{RR}}\alpha_{\mathrm{LL}}-\alpha_{\mathrm{RL}}\alpha_{\mathrm{LR}}}{\alpha_{\mathrm{RR}}-\alpha_{\mathrm{LR}}}\right)\left.\frac{\partial N_{\mathrm{DQD}}}{\partial T_{\mathrm{e}}}\right|_{V_{\mathrm{LP}}}=k_{\mathrm{B}}\left.\frac{\partial N_{\mathrm{DQD}}}{\partial\theta}\right|_{V_{\mathrm{LP}}}. \tag{3}$$

Hence, the entropy map in Fig. 2(c) was obtained from the integral $\Delta S=k_{\mathrm{B}}\int\frac{\partial N}{\partial\theta}dV_{\mathrm{LP}}$ of this Maxwell relation along each $\delta$-axis trajectory. On the other hand, for a $V_{\mathrm{LP}}$-axis trajectory, the Maxwell relation in equation (1) requires the charge information of $\alpha_{\mathrm{LL}}N_{\mathrm{L}}+\alpha_{\mathrm{RL}}N_{\mathrm{R}}$, which cannot be accessed from a single CS, as discussed in the main text. Nevertheless, we combined the evaluated derivative $\frac{\partial N}{\partial\theta}$ with $\Delta S=k_{\mathrm{B}}\int\frac{\partial N}{\partial\theta}dV_{\mathrm{LP}}$ (equivalently, $\frac{\partial S}{\partial V_{\mathrm{LP}}}|_{T_{\mathrm{e}}}=k_{\mathrm{B}}\frac{\partial N}{\partial\theta}|_{V_{\mathrm{LP}}}$) for each $V_{\mathrm{LP}}$-axis trajectory, to mimic an experimental situation where the entropy measurement protocol[5] for a single QD is applied without any modification. The resulting entropy map is shown in Fig. 2(d). We compared the entropy maps with the Gibbs entropy $S=-k_{\mathrm{B}}\sum_i p_i\ln p_i$ obtained from the computed DQD density matrix $\hat{\rho}_{\mathrm{DQD}}$ via $S(V_{\mathrm{LP}},V_{\mathrm{RP}};T_{\mathrm{e}})=-k_{\mathrm{B}}\mathrm{Tr}_{\mathrm{DQD}}\left[\hat{\rho}_{\mathrm{DQD}}\ln\hat{\rho}_{\mathrm{DQD}}\right]$, where $p_i$ is the probability of finding the DQD in an eigenstate $i$. The entropy map in Fig. 2(c), obtained along $\delta$-axis trajectories, is in excellent agreement with the Gibbs entropy, while the entropy map in Fig. 2(d) shows clear deviations from the Gibbs entropy.

The parameters chosen for the simulation results in Fig. 2(c) and 2(d) are as follows: The Hamiltonian parameters are $V_{\mathrm{L0}}=-0.705$ V, $V_{\mathrm{R0}}=-0.586$ V, $E_{\mathrm{C,L}}=E_{\mathrm{C,R}}=3$ meV, $E_{\mathrm{Cm}}=0.3$ meV, $t_{\mathrm{c}}=6.5$ μeV, $\alpha_{\mathrm{RR}}=\alpha_{\mathrm{LL}}=62$ meV/V, $\alpha_{\mathrm{RL}}=27$ meV/V, and $\alpha_{\mathrm{LR}}=28$ meV/V. The parameters of the CS conductance model are $g_{\mathrm{R}}=g_{\mathrm{L}}=0.02\,\frac{\mathrm{e}^2}{h}$, $\gamma_{\mathrm{R}}=\gamma_{\mathrm{L}}=1.5\,\frac{\mathrm{e}^2}{h}/\mathrm{V}$, and $G_0=0.1\,\frac{\mathrm{e}^2}{h}$. The derivative $\frac{\partial N}{\partial\theta}$ at the average temperature $\bar{T}_{\mathrm{e}}=(T_{\mathrm{e,c}}+T_{\mathrm{e,h}})/2$ was obtained from the difference in $N$ between two temperatures $T_{\mathrm{e,c}}=100$ mK and $T_{\mathrm{e,h}}=200$ mK. The integral $\Delta S=k_{\mathrm{B}}\int\frac{\partial N}{\partial\theta}dV_{\mathrm{LP}}$ was computed from the corresponding summation over discrete values of $V_{\mathrm{LP}}$ with the incremental step size $\Delta V_{\mathrm{LP}}=50$ μV.

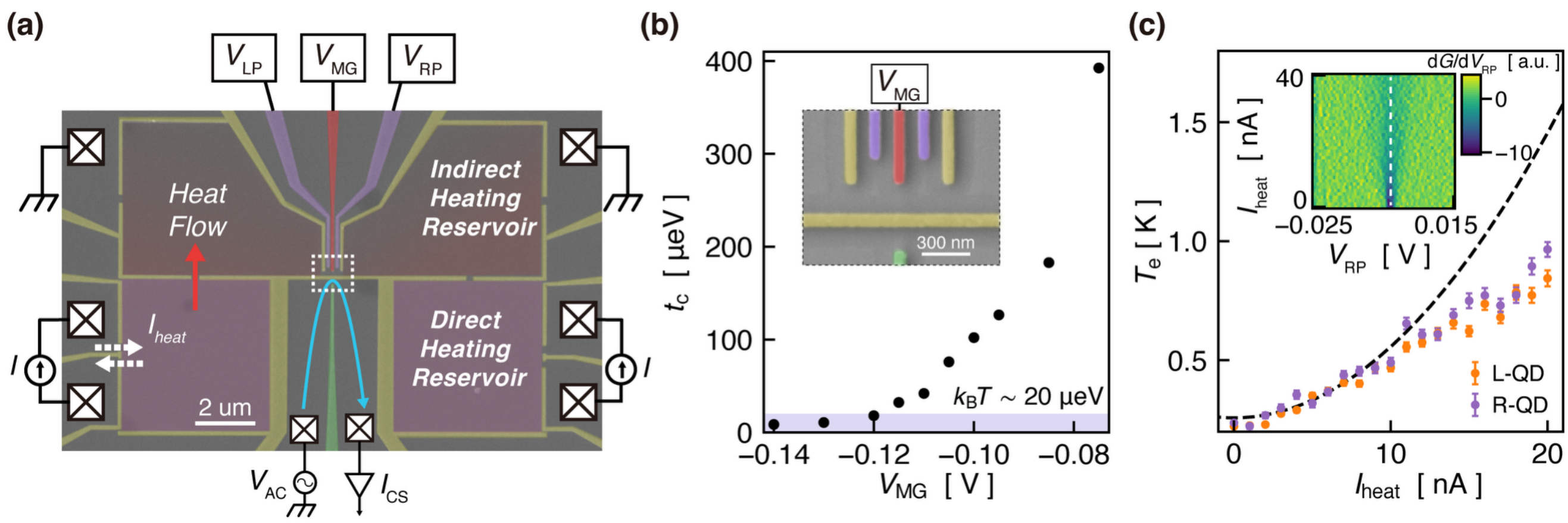

**Figure 1. Device image and calibration. (a)** False-colored scanning electron microscope (SEM) image of the device. The device consists of a serial double quantum dot (DQD) coupled to two independent heating reservoirs and a nearby charge sensor (CS) for occupation readout. The conductance of the CS quantum point contact (QPC), $G = I_{\rm CS}/V_{\rm AC}$, is measured along the blue solid arrow. Each heating reservoir consists of a direct and an indirect heating chamber. The heating current $I_{\rm heat}$ flows through two QPCs in the direct chamber with opposite polarities (white dashed arrow), and the resulting heat propagates into the indirect chamber (red arrow). The gate voltages $V_{\rm LP}$ and $V_{\rm RP}$ control the energy levels of the DQD, while $V_{\rm MG}$ controls the interdot tunnel coupling $t_{\rm c}$. **(b)** Interdot tunnel coupling $t_{\rm c}$ extracted from the CS measurements. The shaded region indicates the thermal energy $k_{\rm B}T_{\rm e} \sim 20$ μeV shown for comparison. The inset shows an enlarged SEM image of the DQD region indicated by the dashed box in (a). **(c)** Electron temperatures of the left and right indirect heating reservoirs, $T_{\rm e,L}$ (orange) and $T_{\rm e,R}$ (purple), measured as functions of the heating current $I_{\rm heat}$. The black dashed curve is a quadratic fit to the data for $I_{\rm heat} < 10$ nA. The inset illustrates $\frac{{\rm d}G}{{\rm d}V_{\rm RP}}$, which was used to extract the electron temperature; the white dashed line indicates the center of the charge transition for $I_{\rm heat} = 0$ nA.

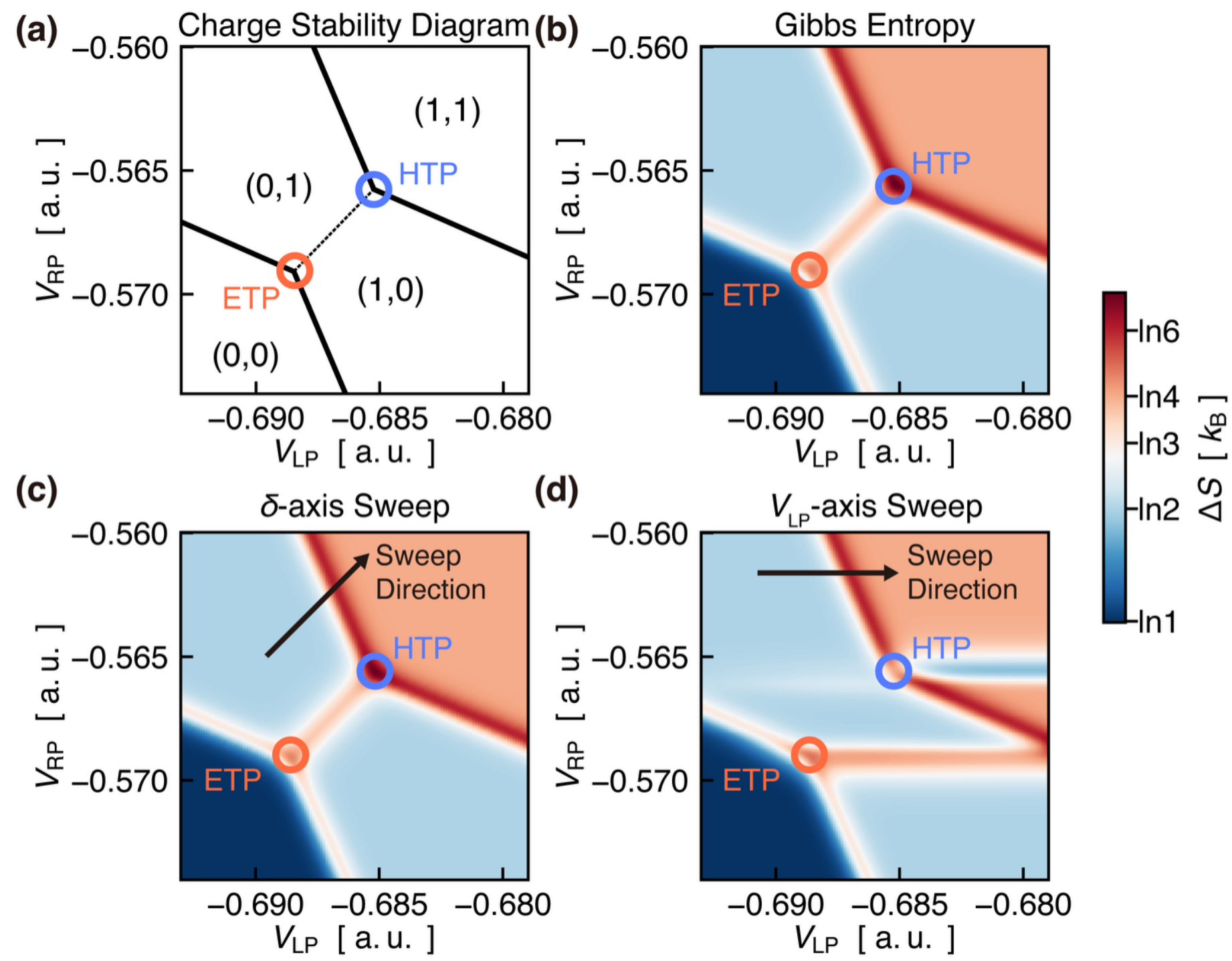

**Figure 2. Charge stability diagram and entropy. (a)** Schematic charge stability diagram of the double quantum dot (DQD). The four charge states $(N_L, N_R) = (0,0)$, $(1,0)$, $(0,1)$, and $(1,1)$ are shown. The electron triple point (ETP) and hole triple point (HTP) are marked. The solid lines indicate the charge transition boundaries; the dashed line marks the boundary between the $(1,0)$ and $(0,1)$ regions. A trajectory parallel to this dashed line corresponds to the $\delta$-axis, along which $\varepsilon = \varepsilon_R - \varepsilon_L$ is kept constant, where $\varepsilon_L$ and $\varepsilon_R$ are the energy levels of the left and right quantum dots (QDs), respectively. **(b)** Theoretical entropy map calculated from the DQD Hamiltonian using the Gibbs entropy formula. **(c)** Theoretical entropy map obtained by applying the Maxwell relation in equation (2) along $\delta$-axis trajectories parallel to the boundary between the $(0,1)$ and $(1,0)$ regions. **(d)** Apparent entropy map obtained by applying the Maxwell relation in equation (1) along trajectories parallel to the $V_{LP}$-axis. The color bar applies to panels (b)–(d). Parameters chosen for the simulations in (b), (c) and (d) are described in Methods.

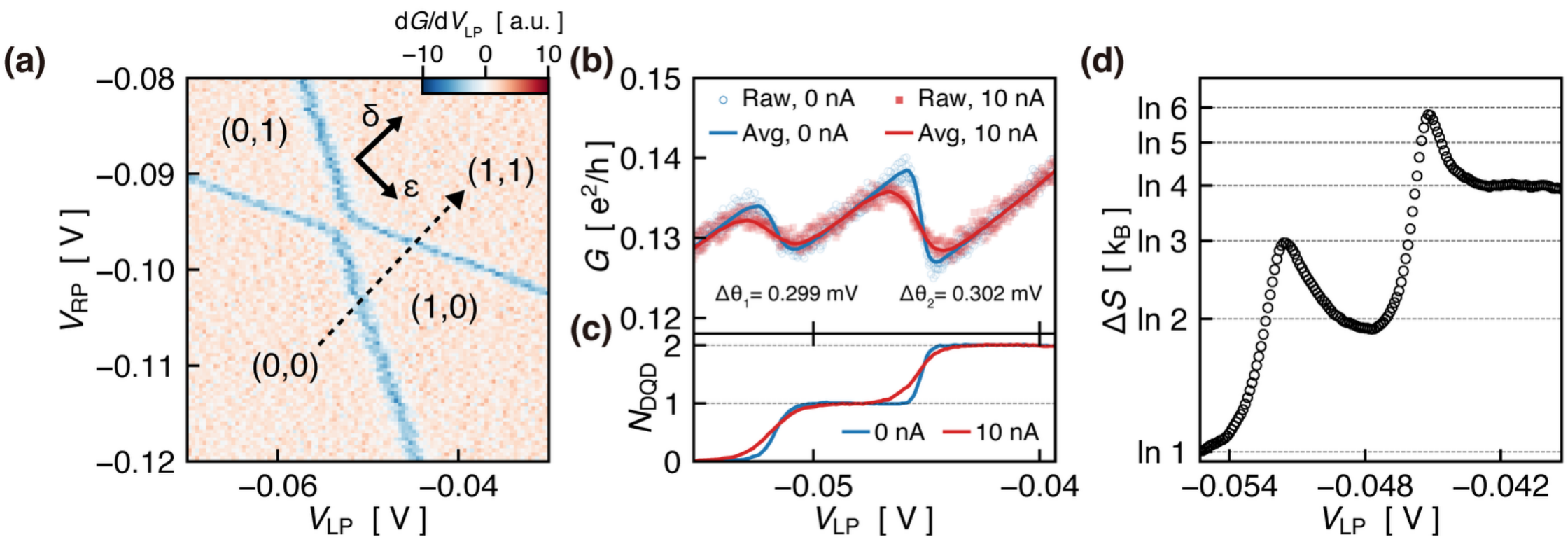

**Figure 3. Measured entropy far away from the triple points. (a)** Charge stability diagram of the double quantum dot (DQD) measured by the charge sensor (CS). The $\varepsilon$- and $\delta$-axes are indicated. The black dashed arrow denotes the entropy measurement path traversing the (0,0) → (1,0) → (1,1) charge regions. **(b)** Conductance measured along the path indicated in (a) for the unheated case (blue circles) and the heated case with an applied current of 10 nA (red squares). The blue and red solid lines represent the averaged conductance. **(c)** Normalized $N_{\mathrm{DQD}}$ along the path indicated in (a) for the unheated (blue) and heated (red) cases. **(d)** Entropy extracted from the $N_{\mathrm{DQD}}$ data shown in (c) via the Maxwell relation.

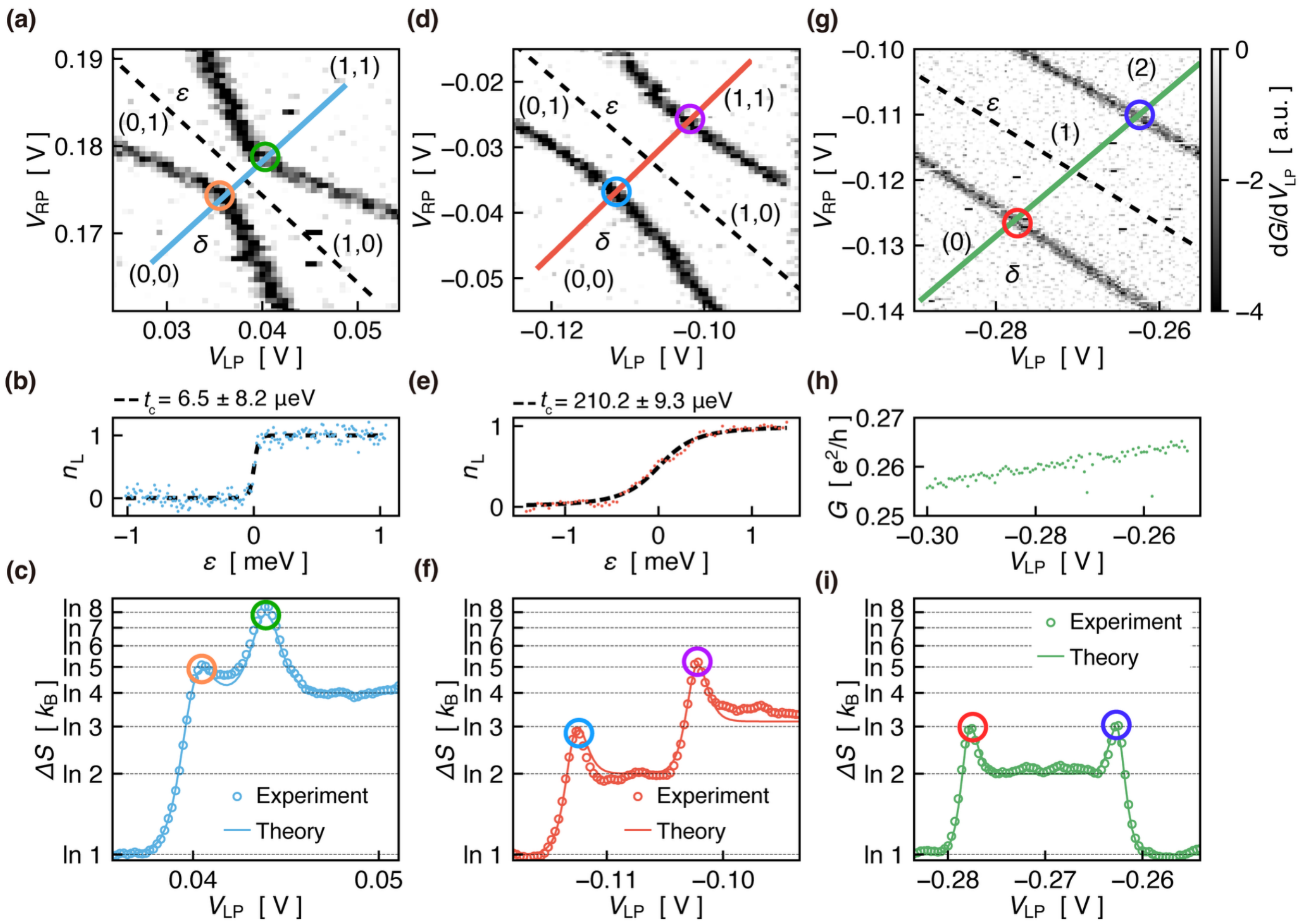

**Figure 4. Measured entropy along the $\delta$-axis with varying interdot tunnel coupling $t_c$. (a)** A DQD charge stability diagram obtained from differentiating the charge sensor (CS) signal in the weak-tunneling regime ($t_c < k_B\bar{T}_e$). The orange and green circles indicate the electron triple point (ETP) and hole triple point (HTP), respectively. The mean population of the left quantum dot (QD), $n_L$, is extracted by measuring CS conductance, $G$, along the $\varepsilon$-axis (black dashed line), while the entropy is measured along the $\delta$-axis (blue solid line). **(b)** Extracted mean population of the left QD along the $\varepsilon$-axis in (a). **(c)** Measured (blue circle markers) and theoretically calculated (blue solid line) entropy of the weak-tunneling regime across the ETP (orange circle) and HTP (green circle). **(d)** A stability diagram for the intermediate-tunneling regime ($t_c > k_B\bar{T}_e$). ETP (HTP) is marked with a cyan (purple) circle. As in (a), the orange solid (black dashed) line indicates the $\delta$-axis ($\varepsilon$-axis). **(e)** Mean population of the left QD along the $\varepsilon$-axis in (d). **(f)** Measured (orange circle markers) and calculated (orange solid line) entropy of the intermediate-tunneling regime across the ETP (cyan circle) and HTP (purple circle). **(g)** A stability diagram for the strong-tunneling regime ($t_c \sim E_C \gg k_B\bar{T}_e$), where the system approaches a merged single QD regime. ETP (HTP) is marked with a red (blue) circle. The $\varepsilon$- and $\delta$-axes are indicated by the black dashed and green solid lines, respectively. **(h)** Measured $G$ along the $\varepsilon$-axis. There is no difference between (1,0) and (0,1) states in $G$, precluding the conversion of $G$ into $n_L$. **(i)** Measured (green circle markers) and calculated (green solid line) entropy of the single QD regime across the ETP (red circle) and HTP (blue circle).

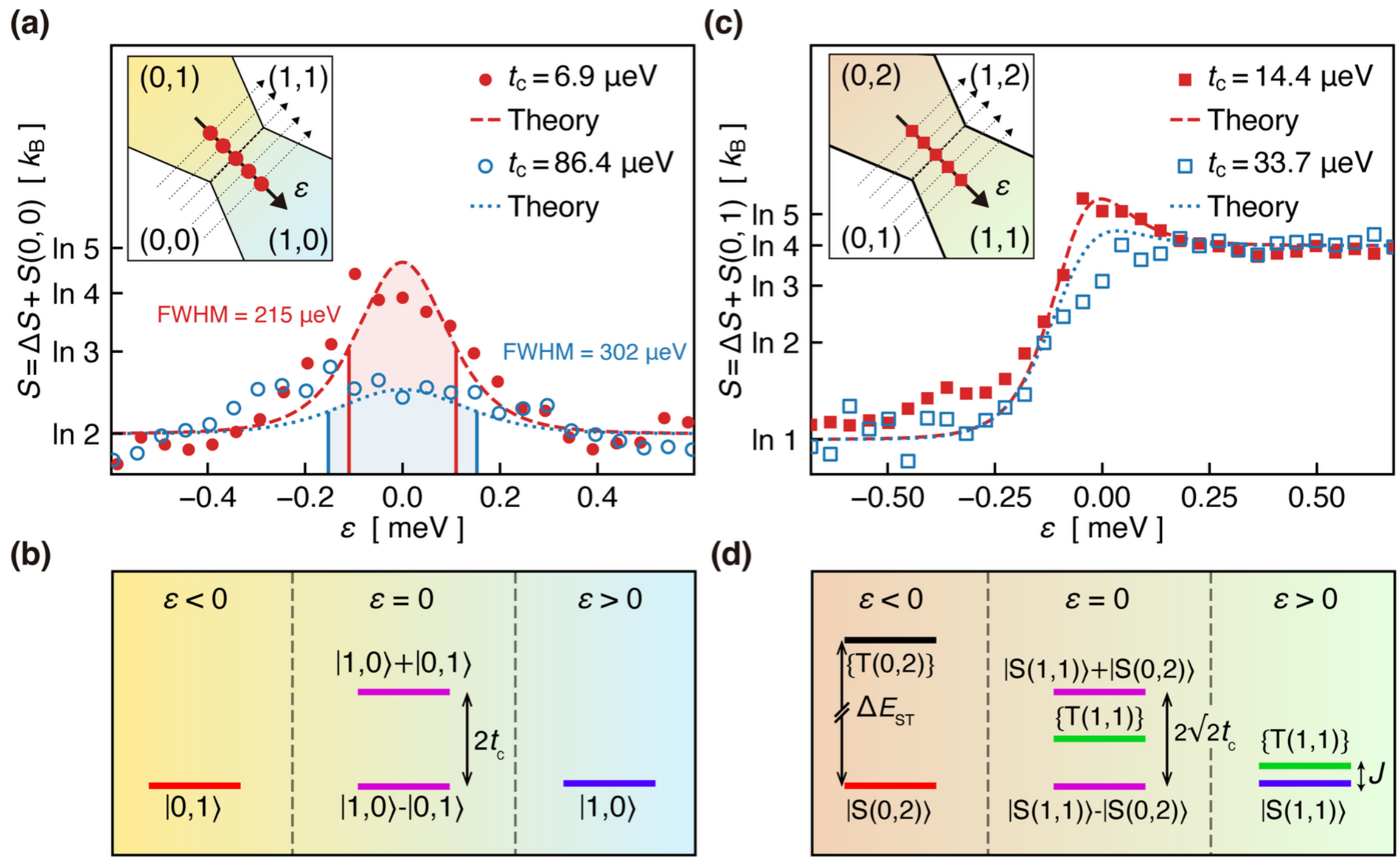

**Figure 5. Entropy evolution across the $\varepsilon$-axis. (a)** Entropy as a function of $\varepsilon$, obtained from entropy traces measured parallel to the $\delta$-axis at various $\varepsilon$. The absolute entropy of $N_{\mathrm{DQD}} = 1$ region, $S(\varepsilon) = \Delta S(\varepsilon) + S(0,0) = \Delta S(\varepsilon)$, is extracted as the minimum in the $N_{\mathrm{DQD}} = 1$ region. The inset shows measurement trajectories from (0,0) to (1,1): the dashed arrows denote each trajectory, and the solid arrow indicates the variable of interest, $\varepsilon$, against which $\Delta S$ is extracted. Red filled circles denote the weak-tunneling regime ($t_c = 6.9$ μeV), and blue open circles denote the intermediate-tunneling regime ($t_c = 86.4$ μeV). The red dashed and the blue dotted lines indicate the theoretically calculated entropy in the weak- and intermediate-tunneling regimes. The shaded region and vertical lines indicate the full width at half maximum (FWHM) of each theoretical curve. **(b)** Schematic energy-level diagram for the $(0,1) \rightarrow (1,0)$ transition. The $|0,1\rangle$ and $|1,0\rangle$ states, as well as the bonding ($|1,0\rangle - |0,1\rangle$) and antibonding states ($|1,0\rangle + |0,1\rangle$), are each doubly degenerate due to spin. The bonding–antibonding energy splitting is $2t_c$ at $\varepsilon = 0$. **(c)** Entropy along the $\varepsilon$-axis obtained in a similar manner for measurements from charge configuration (0,1) to (1,2). Red filled squares denote the weak-tunneling regime ($t_c = 14.4$ μeV), and blue open squares denote the case with a larger interdot tunnel coupling ($t_c = 33.7$ μeV). The red dashed and blue dotted lines indicate the theoretically calculated entropy in the weak- and intermediate-tunneling regimes. Again, $S(\varepsilon)$ denotes the absolute entropy, obtained by adding the entropy of the initial (0,1) state, $S(0,1) = k_B \ln 2$, to the measured entropy change $\Delta S(\varepsilon)$, that is, $S(\varepsilon) = \Delta S(\varepsilon) + k_B \ln 2$. **(d)** Schematic energy-level diagram for the $(0,2) \rightarrow (1,1)$ transition. Inside the (1,1) region, the exchange interaction forms the singlet $|S(1,1)\rangle$ and three degenerate triplet states $\{T(1,1)\} \equiv \{|T_+(1,1)\rangle, |T_0(1,1)\rangle, |T_-(1,1)\rangle\}$ with the exchange energy $J = \frac{4t_c^2}{(E_C - E_{Cm})}$, where $E_C$ is the charging energy of each QD and $E_{Cm}$ is the mutual charging energy. The triplet states in the (0,2) region, $\{T(0,2)\} \equiv \{|T_+(0,2)\rangle, |T_0(0,2)\rangle, |T_-(0,2)\rangle\}$, lie above $|S(0,2)\rangle$ by an energy splitting $\Delta E_{\mathrm{ST}}$, which is much larger than the thermal energy scale $k_B \overline{T}_e$. $|S(1,1)\rangle$ and $|S(0,2)\rangle$ form the bonding singlet state ($|S(1,1)\rangle - |S(0,2)\rangle$) and antibonding singlet state ($|S(1,1)\rangle + |S(0,2)\rangle$), with the energy splitting of $2\sqrt{2}t_c$.

## Supplementary Note 1: Bias Spectroscopy for Determining $\alpha$

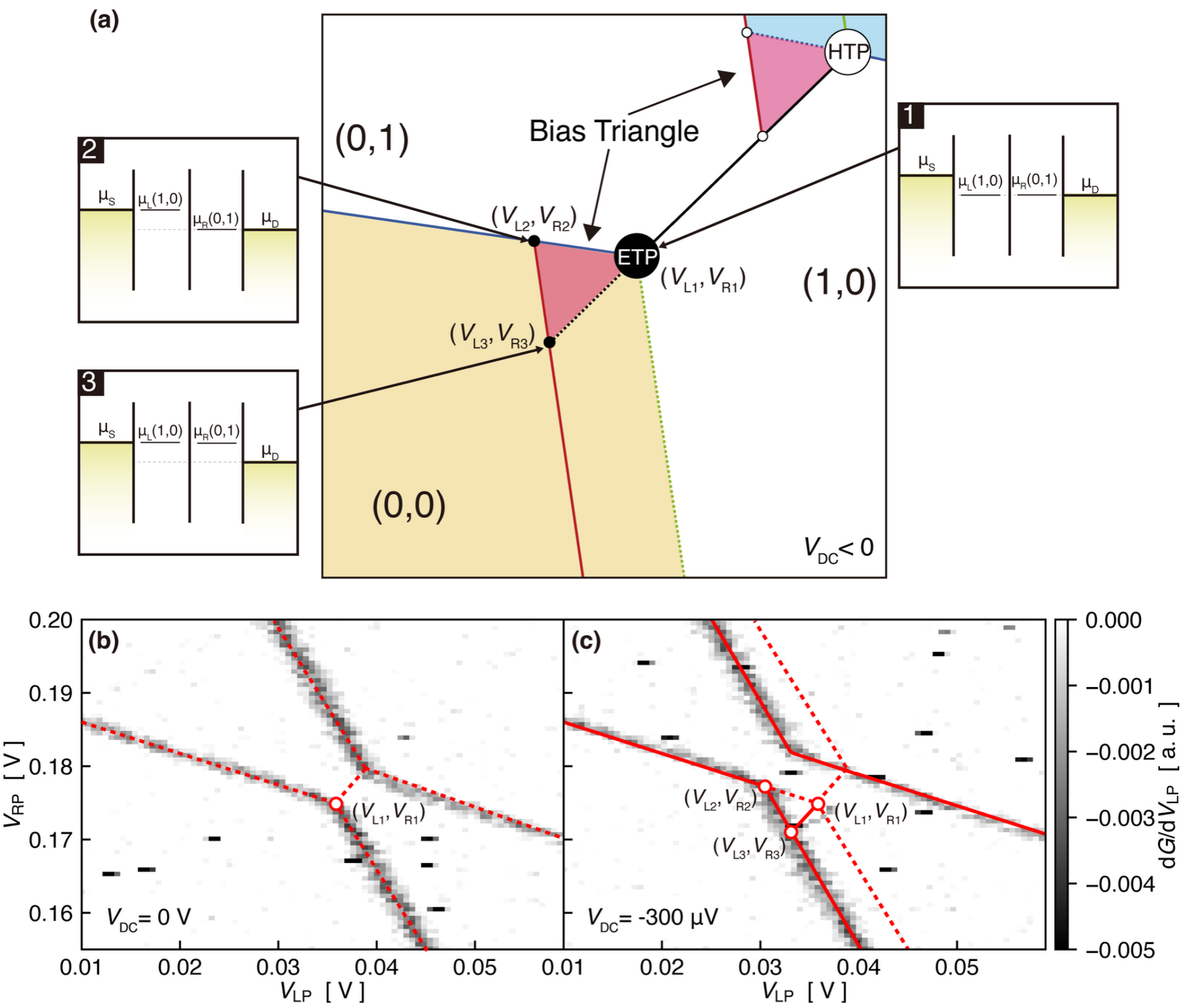

**Supplementary Figure S1. Bias Spectroscopy for Determining the Lever Arm $\alpha$. (a)** Schematic of bias spectroscopy for a double quantum dot (DQD). When a bias ($V_{DC} < 0$) is applied to the left reservoir, the charge transition line of the left quantum dot (L-QD) shifts from its equilibrium position (green dashed line) to a shifted position (red solid line). Under zero-bias conditions, transport is restricted to the electron and hole triple points (ETP and HTP). In contrast, a finite bias allows transport within a triangular region known as the 'bias triangle.' Key coordinates are identified as follows: Point 1 corresponds to the ETP; Point 2 is the intersection of the shifted L-QD transition line and the unperturbed right quantum dot (R-QD) transition line; and Point 3 is the intersection of the shifted L-QD line with the extension of the baseline connecting the ETP and HTP. The corresponding energy-level diagrams for each configuration are shown for better understanding. **(b)** Charge stability diagram under zero-bias conditions, with data corresponding to the main Fig. 4(a–c). The equilibrium charge transition lines and the line connecting the ETP and HTP are indicated by red dashed lines. **(c)** Charge stability diagram with an applied bias of $V_{DC} = -300$ μV. Red solid lines represent the biased charge transition lines. For direct comparison, the reference lines from (b) are superimposed as red dashed lines to highlight the bias-induced shifts. White circles denote the vertices of the bias triangle, along with their corresponding coordinates.

Although the entropy can be fully extracted in voltage units, it is necessary to convert the voltage units into energy to quantitatively compare the results with theoretical calculations. For this reason, we extract the gate lever arm $\alpha$ from bias spectroscopy. In a double quantum dot (DQD), each quantum dot (QD) has its own plunger gate, which primarily changes the energy levels of the corresponding dot. However, due to the proximity of the two dots, crosstalk also exists between a given QD and the opposite plunger gate. To account for this crosstalk, we define the lever-arm matrix as $\alpha_{ij} \equiv -\partial\varepsilon_i/\partial V_j$ ($i, j \in \{L, R\}$), where $i$ and $j$ denote the QD and gate indices, respectively.

Figure S1(a) illustrates the schematic of bias spectroscopy for the DQD. Under zero-bias conditions, the DQD exhibits stable charge states, and transport is restricted to the triple points. As a finite bias is applied, these triple points expand into characteristic 'bias triangles' due to the shifting of charge transition lines and the opening of a transport window. By analyzing the dimensions and orientation of these bias triangles, we can experimentally determine all components of the lever-arm matrix $\alpha_{ij}$.

To determine the lever-arm matrix, we identify three characteristic points in the bias triangle:

- Point 1 $(V_{\mathrm{L1}}, V_{\mathrm{R1}})$: The voltage coordinates of the electron triple point (ETP).
- Point 2 $(V_{\mathrm{L2}}, V_{\mathrm{R2}})$: The position where the energy level $\mu_{\mathrm{L}}$ of the left QD (L-QD) aligns with the biased chemical potential of the left reservoir ($\mu_L = -eV_{\mathrm{DC}}$).
- Point 3 $(V_{\mathrm{L3}}, V_{\mathrm{R3}})$: The configuration where the energy levels of both QDs are aligned with the chemical potential of the left reservoir.

Based on these definitions, the energy level shifts at each point can be described by the following system of linear equations[1] :

$$\begin{pmatrix} V_{\mathrm{L3}} - V_{\mathrm{L1}} & 0 & V_{\mathrm{R3}} - V_{\mathrm{R1}} & 0 \\ 0 & V_{\mathrm{L3}} - V_{\mathrm{L1}} & 0 & V_{R3} - V_{\mathrm{R1}} \\ V_{\mathrm{L2}} - V_{\mathrm{L1}} & 0 & V_{\mathrm{R2}} - V_{\mathrm{R1}} & 0 \\ 0 & V_{\mathrm{L2}} - V_{\mathrm{L1}} & 0 & V_{\mathrm{R2}} - V_{\mathrm{R1}} \end{pmatrix} \begin{pmatrix} \alpha_{\mathrm{LL}} \\ \alpha_{\mathrm{RL}} \\ \alpha_{\mathrm{LR}} \\ \alpha_{\mathrm{RR}} \end{pmatrix} = \begin{pmatrix} eV_{\mathrm{DC}} \\ eV_{\mathrm{DC}} \\ eV_{\mathrm{DC}} \\ 0 \end{pmatrix} \tag{S1}$$

Figure S1(b) and (c) present representative charge stability diagrams for the bias spectroscopy measurements shown in Fig. 4(a–c) of the main text, comparing the system behavior with and without a DC bias of $V_{\mathrm{DC}} = -300$ μV applied to the left reservoir. In the zero-bias case (Fig. S1(b)), the red dashed line indicates the charge transition line. This reference line is also superimposed in Fig. S1(c) to highlight the shift in the energy level of the L-QD induced by the applied bias. The resulting shifted transition line is represented by the red solid line. By comparing these two configurations, we can precisely identify the vertices of the 'bias triangle' as labeled in (c). Finally, by solving the system of linear equations in equation (S1), we determine all components of the lever-arm matrix $\alpha_{ij}$.

With the lever-arm matrix $\alpha_{ij}$ determined, we can derive the effective lever arms along the $\delta$-axis and $\varepsilon$-axis, noting that along the $\delta$-axis $d\delta = d\mu_{\mathrm{L}} = d\mu_{\mathrm{R}}$ holds and along the $\varepsilon$-axis, $d\varepsilon = d\mu_{\mathrm{R}} - d\mu_{\mathrm{L}} = -2d\mu_{\mathrm{L}}$. From simple algebra, we can derive

$$\alpha_\delta = \frac{\alpha_{\mathrm{LL}}\alpha_{\mathrm{RR}} - \alpha_{\mathrm{RL}}\alpha_{\mathrm{LR}}}{\alpha_{\mathrm{RR}} - \alpha_{\mathrm{LR}}}, \tag{S2}$$

and

$$\alpha_\varepsilon = \frac{2(\alpha_{\mathrm{LL}}\alpha_{\mathrm{RR}} - \alpha_{\mathrm{RL}}\alpha_{\mathrm{LR}})}{\alpha_{\mathrm{RR}} + \alpha_{\mathrm{LR}}}. \tag{S3}$$

**Table S1. Lever arms for reported data.**

| Used in Figures | $\lvert\alpha_{\mathrm{LL}}\rvert$ [meV/V] | $\lvert\alpha_{\mathrm{RL}}\rvert$ [meV/V] | $\lvert\alpha_{\mathrm{LR}}\rvert$ [meV/V] | $\lvert\alpha_{\mathrm{RR}}\rvert$ [meV/V] | $\lvert\alpha_{\delta}\rvert$ [meV/V] | $\lvert\alpha_{\varepsilon}\rvert$ [meV/V] |
|---|---|---|---|---|---|---|
| 4(a–c) | 62 | 27 | 28 | 62 | 91 | 69 |
| 4(d–f) | 54 | 36 | 34 | 56 | 82 | 40 |
| 5(a) – Red Circle | 63 | 28 | 28 | 63 | 91 | 70 |
| 5(a) – Blue Circle | 74 | 41 | 40 | 76 | 111 | 69 |
| 5(c) – Red Square | 57 | 24 | 25 | 56 | 84 | 64 |
| 5(c) – Blue Square | 62 | 28 | 30 | 60 | 96 | 64 |

Table S1 shows the calculated lever-arm values used in the main text. For the case where the two QDs are merged into a single QD (SQD) (Fig. 4(g–i)), we cannot obtain the lever-arm matrix, as the bias triangle is not formed.

Another point that we can examine with the bias spectroscopy is the excited states of the DQD. When the chemical potential of the reservoir is sufficiently elevated to align with an excited level, that state becomes accessible for transport or occupation, which is detected by the CS. By applying the previously determined lever-arm matrix $\alpha_{ij}$, we can quantitatively extract the excitation energy, $\Delta E \equiv E_{\mathrm{es}} - E_{\mathrm{gs}}$, where $E_{\mathrm{es}}$ and $E_{\mathrm{gs}}$ correspond to the excited-state and ground-state energies, respectively. In our study, signatures of these excited states were exclusively observed in the bias spectroscopy data corresponding to main Fig. 4(a–c); no other measurements exhibited similar features within the applied bias range. For the data in Fig. 4(a–c), the extracted excitation energy is $\Delta E = 300$ μeV, which is approximately an order of magnitude larger than the average thermal energy scale of $k_{\mathrm{B}}\overline{T}_{\mathrm{e}} \sim 40$ μeV. Therefore, we can safely neglect the thermal occupation of excited states in our analysis.

## Supplementary Note 2: Validation of Charge Sensor Thermometry via Coulomb Blockade Thermometry

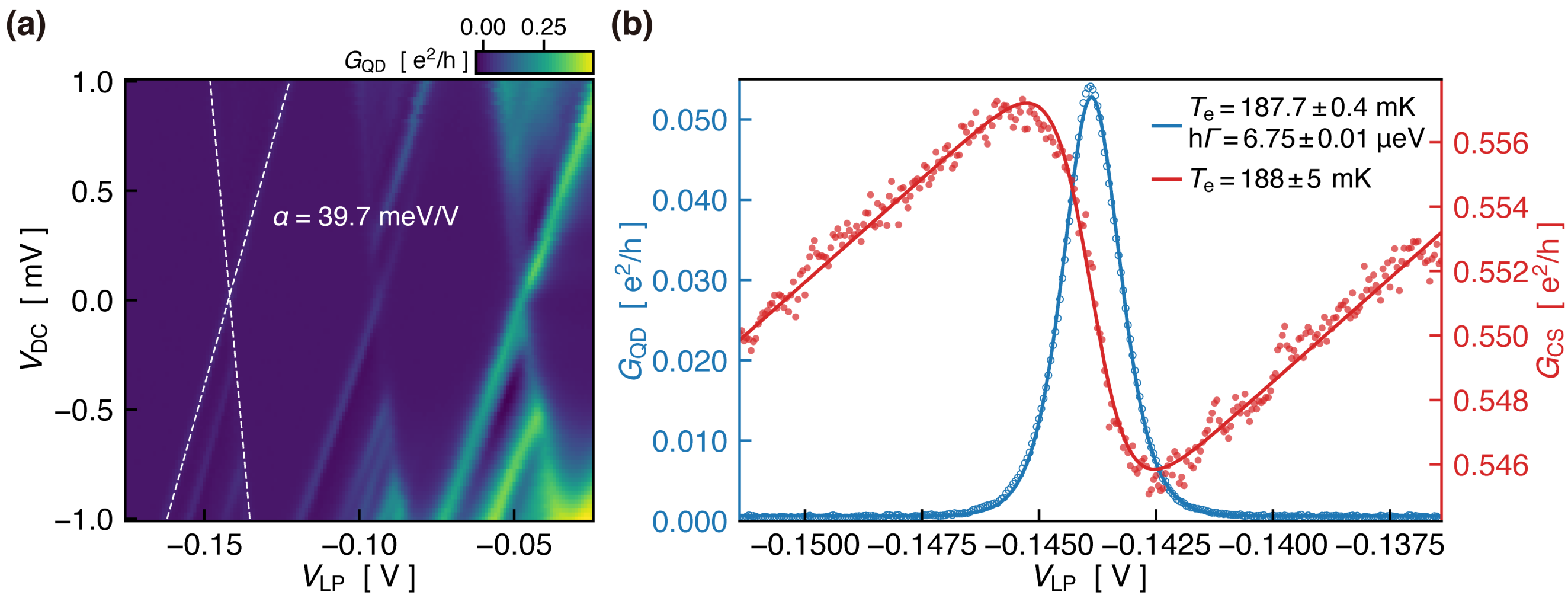

**Supplementary Figure S2. Comparison of Electron Temperatures Obtained from Coulomb Blockade and Charge Sensor.** **(a)** Coulomb diamond measurement of the left quantum dot (L-QD) obtained from direct transport. All gates defining the right quantum dot (R-QD) are positively biased such that the R-QD is not formed. The white dashed lines indicate the diamond edges used to extract the lever arm, giving $\alpha = 39.7$ meV/V. **(b)** Comparison of thermometry based on direct transport and charge sensing. The direct transport conductance $G_{\mathrm{QD}}$ (blue open circles and solid line) and the charge sensor conductance $G_{\mathrm{CS}}$ (red filled circles and solid line) are fitted using the thermal broadening models described in the text.

In the main text, the electron temperature $T_{\mathrm{e}}$ was determined as $T_{\mathrm{e}} = \frac{\alpha\theta}{k_{\mathrm{B}}}$, where $\alpha$ is the lever arm, $\theta$ is the thermal broadening of the charge sensor (CS) signal expressed in voltage units, and $k_{\mathrm{B}}$ is the Boltzmann constant. To verify the validity of this method, we compared the temperature extracted from the thermal broadening of the CS with the temperature independently obtained from Coulomb blockade thermometry[2–4].

To this end, we first formed only the L-QD and simultaneously measured the direct transport across the QD ($G_{\mathrm{QD}}$) and the conductance of the CS ($G_{\mathrm{CS}}$). Figure S2(a) shows the Coulomb diamond[3] obtained from the direct transport measurement. From the slopes of the diamond edges, we determined the lever arm to be $\alpha = 39.7$ meV/V. Using this value of $\alpha$, we extracted $T_{\mathrm{e}}$ from both measurement methods, as shown in Fig. S2(b). The direct transport conductance (blue open circles and line) was fitted using[2] $G_{\mathrm{QD}} = \frac{e^2}{4k_{\mathrm{B}}T_{\mathrm{e}}}\frac{\Gamma}{2}\mathrm{sech}^2\left(\frac{\alpha(V-V_0)}{2k_{\mathrm{B}}T_{\mathrm{e}}}\right)$ where $\Gamma$ is the tunnel rate and $V_0$ is the center of the peak, assuming symmetric coupling to the reservoir[4]. $G_{\mathrm{CS}}$ (red filled circles and line) was fitted to $G_{\mathrm{CS}} = -\frac{I_0}{2}\tanh\left(\frac{\alpha(V-V_0)}{2k_{\mathrm{B}}T_{\mathrm{e}}}\right) + \gamma_1 V + \gamma_2$, where $I_0$ is the sensor conductance step height, and $\gamma_1$ and $\gamma_2$ account for the linear background contribution from the plunger gate of the QD [5,6]. The values of $T_{\mathrm{e}}$ obtained from the two fitting procedures agree well with each other, demonstrating that thermometry based solely on the CS is reliable as long as the lever arm $\alpha$ is accurately determined.

Another point we examined is the magnitude of the reservoir–QD tunnel coupling $\Gamma$. Using the fitting form of $G_{\mathrm{QD}}$ introduced above and assuming symmetric tunnel couplings to the source and drain, we extracted a tunnel coupling of $h\Gamma = 6.75$ μeV, which is approximately one third of the electron temperature in energy units.

In contrast, in the entropy measurements presented in the main text, the device was operated in a regime where no Coulomb blockade peaks were visible in direct transport. In this regime, we can place an upper bound on $\Gamma$ by noting that the conductance noise floor of our experimental setup is $\sim 0.001\ e^2/h$ and that $G_{\mathrm{QD}}$ is directly proportional to $\Gamma$ in the weak-coupling limit. From this consideration, we can indirectly estimate that the effective tunnel coupling in the actual measurement configuration is on the order of several tens of neV, which is much smaller than the relevant thermal energy scale, on the order of several tens of μeV.

## Supplementary Note 3: Process for Obtaining the Interdot Tunnel Coupling $t_c$

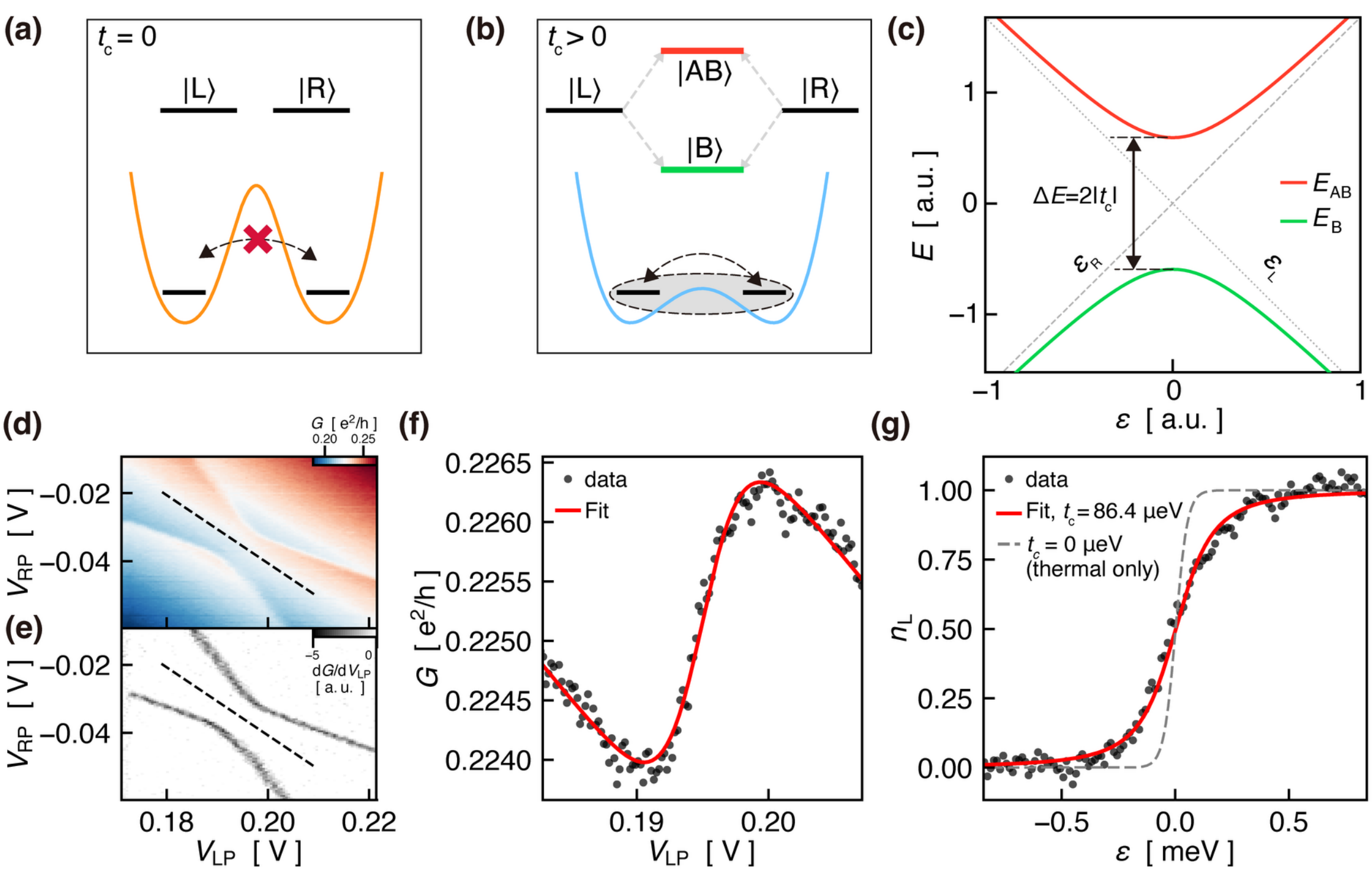

**Supplementary Figure S3. Interdot Tunnel Coupling and the Formation of Bonding–antibonding Molecular States. (a)** Schematic potential and energy-level diagram for $t_c = 0$, where the localized states $|L\rangle$ and $|R\rangle$ remain independent and tunneling is forbidden. **(b)** Schematic potential and energy-level diagram for $t_c > 0$, where the two localized states hybridize into a low-energy bonding state $|B\rangle$ and a high-energy antibonding state $|AB\rangle$. **(c)** Energy levels $E_{AB}$ and $E_B$ as a function of detuning ε, calculated by diagonalizing the two-level Hamiltonian (equation (S4)), with the minimum energy splitting $\Delta E = 2|t_c|$ at $\varepsilon = 0$. **(d)** Charge stability diagram obtained from the CS conductance $G$ as a function of $V_{LP}$ and $V_{RP}$. **(e)** Numerical derivative $dG/dV_{LP}$ of the data in (d). The measurement path for the latter data is indicated with the dashed line for (d) and (e). **(f)** $G$ measured along the ε-axis trace indicated by the dashed line in (d) and (e) (black circles), together with a fit (red line). **(g)** Extracted mean population $n_L$ as a function of detuning ε (black circles), with a fit yielding $t_c = 86.4$ μeV (red line). To more directly visualize the broadening induced by hybridization, we also include the mean population expected from thermal broadening alone ($t_c = 0$), shown as the gray dashed line.

In a DQD, the interdot tunnel coupling $t_c$ quantifies the coupling between electrons that occupy the L-QD and R-QD. Consider the case when $N_{DQD} = 1$ and we vary the energy of the DQD along the ε-axis defined in the main text. When the potential barrier between the two QDs is too high for an electron to tunnel through ($t_c = 0$), tunneling is forbidden, and the energy levels of each QD do not hybridize, remaining as independent energy levels (Fig. S3(a)). Once the potential barrier becomes low enough, tunneling can occur, and the two localized charge states are coupled via the interdot tunnel coupling $t_c$ (Fig. S3(b)). Along the ε-axis, the Hamiltonian of the DQD can be written as[7]

$$H = \begin{pmatrix} \frac{\varepsilon}{2} & t_c \\ t_c & -\frac{\varepsilon}{2} \end{pmatrix} \tag{S4}$$

in the basis of $\{|n_R\rangle, |n_L\rangle\}^T$, where $|n_i\rangle$ denotes the number state. Here, we exclude the spin degeneracy for simplicity; it can be included afterward since spin-flip scattering between the QDs is not considered. Diagonalizing this Hamiltonian yields a low-energy state, denoted bonding (B), and a high-energy state, denoted antibonding (AB). The resulting energy levels as a function of $\varepsilon$ are plotted in Fig. S3(c), where the minimum energy splitting between the B and AB states occurs at the charge-degeneracy point ($\varepsilon = 0$) and is given by $2|t_\mathrm{c}|$. By convolving the Fermi distribution $f(E, T_\mathrm{e})$ and the population of the L-QD $\langle n_\mathrm{L} \rangle$, we obtain the mean population along the ε-axis $n_\mathrm{L}(\varepsilon)$ as [7,8]

$$n_\mathrm{L}(\varepsilon) = \frac{1}{2}\left[1 + \frac{\varepsilon}{\sqrt{\varepsilon^2 + 4t_\mathrm{c}^2}} \tanh\left(\frac{\sqrt{\varepsilon^2 + 4t_\mathrm{c}^2}}{2k_\mathrm{B}T_\mathrm{e}}\right)\right] \tag{S5}$$

To extract $t_\mathrm{c}$ from measurement, we fit the mean population $n_\mathrm{L}$ (or equivalently $n_\mathrm{R}$) as a function of $\varepsilon$ to equation (S5). Figure S3(d) shows the measured charge stability diagram, and Fig. S3(e) shows its numerical derivative with respect to $V_\mathrm{LP}$, taken for better visibility of the charge transition lines. The line perpendicular to the line connecting the ETP and HTP is used as the $\varepsilon$-axis, and the conductance is measured along this axis. Because our CS has a slightly different sensitivity to the L-QD and R-QD, the conductance differs slightly between the (1,0) and (0,1) states (Fig. S3(f)). We subtract the linear background and normalize the resulting signal between 0 and 1, yielding $n_\mathrm{L}(\varepsilon)$ as shown in Fig. S3(g). This fit requires the electron temperature to be determined beforehand, following the procedure detailed in Supplementary Note 2. The voltage-to-energy conversion along the $\varepsilon$-axis is described in Supplementary Note 1.

## Supplementary Note 4: Compensation of the Current-induced Bias

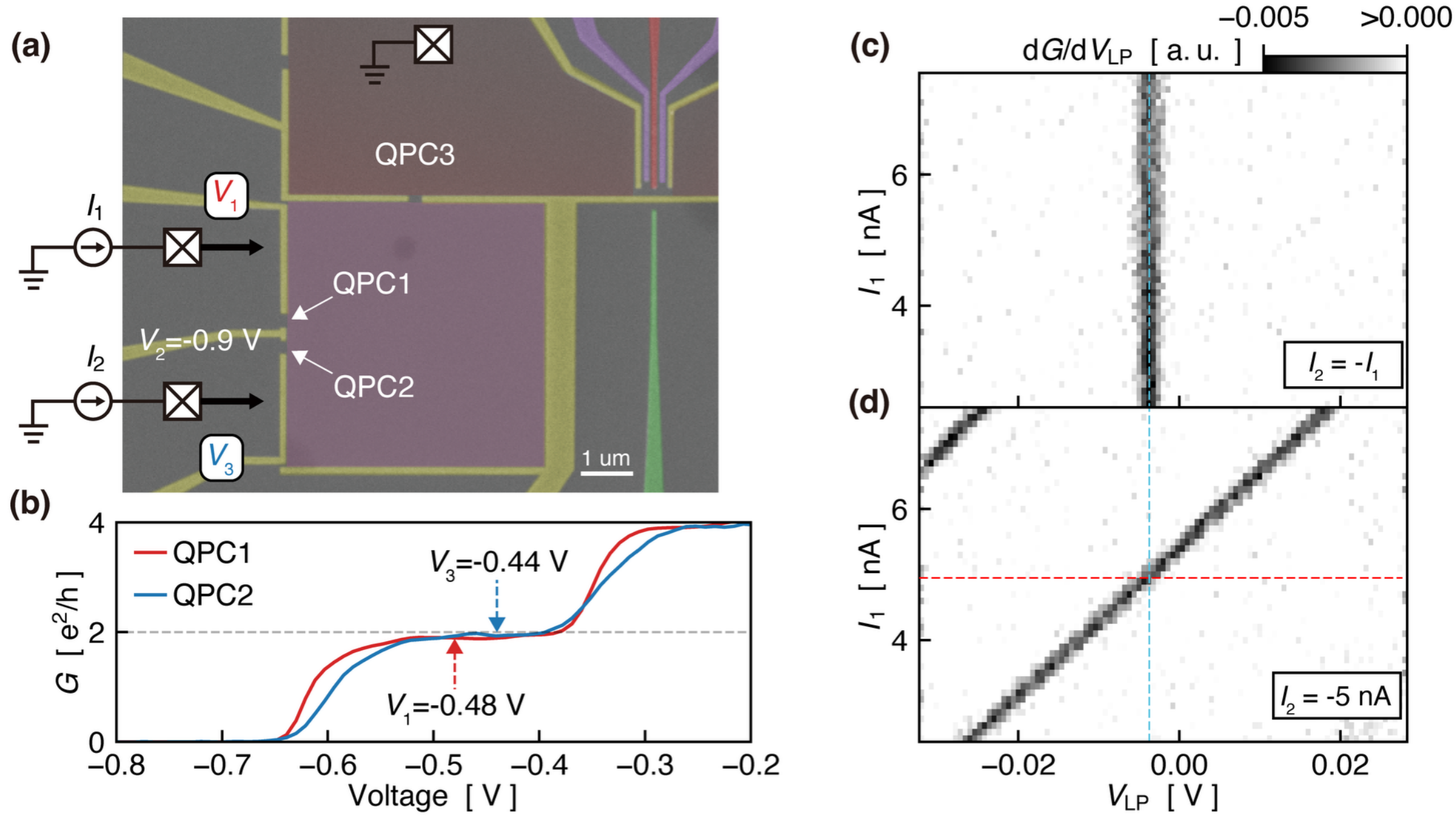

**Supplementary Figure S4. Bias Compensation of the Heating Current. (a)** A scanning electron microscope (SEM) image of the direct heating reservoir. The direct heating reservoir consists of three quantum point contacts (QPCs): QPC1 and QPC2 for Joule heating current and QPC3 for connecting to the indirect heating reservoir. The heating currents $I_1$ and $I_2$ are passed through QPC1 and QPC2, respectively. **(b)** Conductance of QPC1 (red) and QPC2 (blue) as a function of the gate voltage. The two QPCs are defined by a shared central gate ($V_2 = -0.9$ V, fixed), while their individual conductances are independently measured. The control voltages $V_1$ and $V_3$ are adjusted to place each QPC near the midpoint of the first conductance plateau ($G \approx 2e^2/h$), as indicated by the arrows with their corresponding colors. **(c)** Heating result with the compensation method measured using the charge sensor (CS). Here, $I_2 = -I_1$. **(d)** Heating result without compensation. $I_2$ was fixed at $-5$ nA while $I_1$ was varied. The dashed vertical line in (c) and (d) indicates the original position of the QD energy level. The horizontal dashed line in (d) marks the condition $I_2 = -I_1$.

Heating of the reservoir was achieved by driving currents through two separate quantum point contacts (QPCs), labeled QPC1 and QPC2, as shown in Fig. S4(a). Each QPC was biased using an independent current source made by a voltage source in series with a 1 MΩ resistor, supplying currents $I_1$ and $I_2$, respectively. Both QPC1 and QPC2 were placed in the middle of the first conductance plateau, as shown in Fig. S4(b).

As discussed in the main text, the entropy measurement protocol based on the Maxwell relation requires that all reservoirs remain in thermal equilibrium. To minimize the current-induced bias, we therefore applied currents to QPC1 and QPC2 with opposite polarities, i.e., $I_1 = -I_2$. Figure S4(c) shows the resulting heating generated by this compensation scheme. Importantly, the CS signal exhibits a negligible shift under this condition (light-blue dashed line), indicating that our heating method does not induce a significant bias in the system.

The presence of a current-induced bias becomes evident when the compensation scheme is not applied. Figure S4(d) shows the CS response measured while fixing $I_2 = -5$ nA and sweeping $I_1$. As can be seen, in the absence of current compensation the QD energy levels exhibit a substantial shift, whereas the original peak position is recovered only at the compensated condition $I_1 = -I_2$. This observation provides further evidence that our heating method does not induce a current-induced bias in the system.

## Supplementary Note 5: Derivation of Maxwell Relations in Equations (1) and (2)

We derive the Maxwell relations for $V_{\mathrm{LP}}$-axis and $\delta$-axis trajectories, presented as equations (1) and (2) in the main text, respectively. The differential grand potential of the DQD system[9–11] is written as

$$d\Phi = -S dT_{\mathrm{e}} + N_{\mathrm{L}} d\varepsilon_{\mathrm{L}} + N_{\mathrm{R}} d\varepsilon_{\mathrm{R}} + \cdots, \tag{S6}$$

where $S$ is the total entropy of the system, $T_{\mathrm{e}}$ is the electron temperature, $N_L$ and $N_R$ are the average occupations of the L-QD and R-QD, respectively, and $\varepsilon_{\mathrm{L}}$ and $\varepsilon_{\mathrm{R}}$ are their energy levels. Since the electron temperatures of the left and right reservoirs, $T_{\mathrm{e,L}}$ and $T_{\mathrm{e,R}}$, are approximately equal in our measurements, we set $T_{\mathrm{e,L}} \simeq T_{\mathrm{e,R}} \equiv T_{\mathrm{e}}$. The terms associated with parameters that are held fixed during the measurement are omitted.

The QD energy levels are controlled by the two plunger gate voltages through the lever-arm coefficients $\alpha_{ij} \equiv -\partial \varepsilon_i / \partial V_j$ $(i, j \in \{L, R\})$, where $i$ and $j$ denote the QD and gate indices, respectively (Supplementary Note 1). Accordingly, their variations are $d\varepsilon_{\mathrm{L}} = -\alpha_{\mathrm{LL}} dV_{\mathrm{LP}} - \alpha_{\mathrm{LR}} dV_{\mathrm{RP}}$ and $d\varepsilon_{\mathrm{R}} = -\alpha_{\mathrm{RL}} dV_{\mathrm{LP}} - \alpha_{\mathrm{RR}} dV_{\mathrm{RP}}$. Substituting these expressions into equation (S6) yields

$$d\Phi = -S dT_{\mathrm{e}} - (\alpha_{\mathrm{LL}} N_{\mathrm{L}} + \alpha_{\mathrm{RL}} N_{\mathrm{R}}) dV_{\mathrm{LP}} - (\alpha_{\mathrm{LR}} N_{\mathrm{L}} + \alpha_{\mathrm{RR}} N_{\mathrm{R}}) dV_{\mathrm{RP}} + \cdots. \tag{S7}$$

For a $V_{\mathrm{LP}}$-axis trajectory, $V_{\mathrm{LP}}$ is varied while $V_{\mathrm{RP}}$ is fixed. From equation (S7), the first derivatives of the grand potential with respect to $T_{\mathrm{e}}$ and $V_{\mathrm{LP}}$ are $\frac{\partial \Phi}{\partial T_{\mathrm{e}}}\Big|_{V_{\mathrm{LP}}, V_{\mathrm{RP}}} = -S$, and $\frac{\partial \Phi}{\partial V_{\mathrm{LP}}}\Big|_{T_{\mathrm{e}}, V_{\mathrm{RP}}} = -(\alpha_{\mathrm{LL}} N_{\mathrm{L}} + \alpha_{\mathrm{RL}} N_{\mathrm{R}})$. Taking the mixed second derivatives of $\Phi$ with respect to $T_{\mathrm{e}}$ and $V_{\mathrm{LP}}$, and using $\frac{\partial^2 \Phi}{\partial V_{\mathrm{LP}} \partial T_{\mathrm{e}}} = \frac{\partial^2 \Phi}{\partial T_{\mathrm{e}} \partial V_{\mathrm{LP}}}$, we obtain

$$\frac{\partial S}{\partial V_{\mathrm{LP}}}\bigg|_{T_{\mathrm{e}}, V_{\mathrm{RP}}} = \frac{\partial (\alpha_{\mathrm{LL}} N_{\mathrm{L}} + \alpha_{\mathrm{RL}} N_{\mathrm{R}})}{\partial T_{\mathrm{e}}}\bigg|_{V_{\mathrm{LP}}, V_{\mathrm{RP}}} \tag{S8}$$

which is equation (1) in the main text.

In contrast, along the $\delta$-axis, the two QD energy levels are shifted by the same amount $d\delta = d\varepsilon_{\mathrm{L}} = d\varepsilon_{\mathrm{R}}$, so that the energy difference $\varepsilon = \varepsilon_{\mathrm{R}} - \varepsilon_{\mathrm{L}}$ remains constant. Equation (S6) then reduces to $d\Phi = -S dT_{\mathrm{e}} + N_{\mathrm{DQD}} d\delta + \cdots$ in terms of the total occupation $N_{\mathrm{DQD}} = N_{\mathrm{L}} + N_{\mathrm{R}}$ of the DQD. Applying the same argument to the mixed derivatives with respect to $T_{\mathrm{e}}$ and $\delta$ leads to

$$\frac{\partial S}{\partial \delta}\bigg|_{T_{\mathrm{e}}, \varepsilon} = -\frac{\partial N_{\mathrm{DQD}}}{\partial T_{\mathrm{e}}}\bigg|_{\delta, \varepsilon} \tag{S9}$$

which is equation (2) in the main text. Experimentally, the $\delta$-axis condition is implemented by varying the two plunger gate voltages simultaneously such that $d\varepsilon_{\mathrm{L}} = d\varepsilon_{\mathrm{R}}$, giving $dV_{\mathrm{RP}} = \frac{\alpha_{\mathrm{LL}} - \alpha_{\mathrm{RL}}}{\alpha_{\mathrm{RR}} - \alpha_{\mathrm{LR}}} dV_{\mathrm{LP}}$. Together with the fixed energy difference $\varepsilon$, the $\delta$-axis trajectory in the plunger gate space is defined as $V_{\mathrm{RP}} = \frac{\alpha_{\mathrm{LL}} - \alpha_{\mathrm{RL}}}{\alpha_{\mathrm{RR}} - \alpha_{\mathrm{LR}}} V_{\mathrm{LP}} - \frac{\varepsilon}{\alpha_{\mathrm{RR}} - \alpha_{\mathrm{LR}}}$. Along the $\delta$-axis, $d\delta$ can be expressed in terms of $dV_{\mathrm{LP}}$ as $d\delta = -\alpha_{\delta} dV_{\mathrm{LP}}$, where $\alpha_{\delta} \equiv \frac{\alpha_{\mathrm{RR}} \alpha_{\mathrm{LL}} - \alpha_{\mathrm{RL}} \alpha_{\mathrm{LR}}}{\alpha_{\mathrm{RR}} - \alpha_{\mathrm{LR}}}$ is the effective lever arm.

## Supplementary Note 6: Data Processing and Entropy Extraction Process

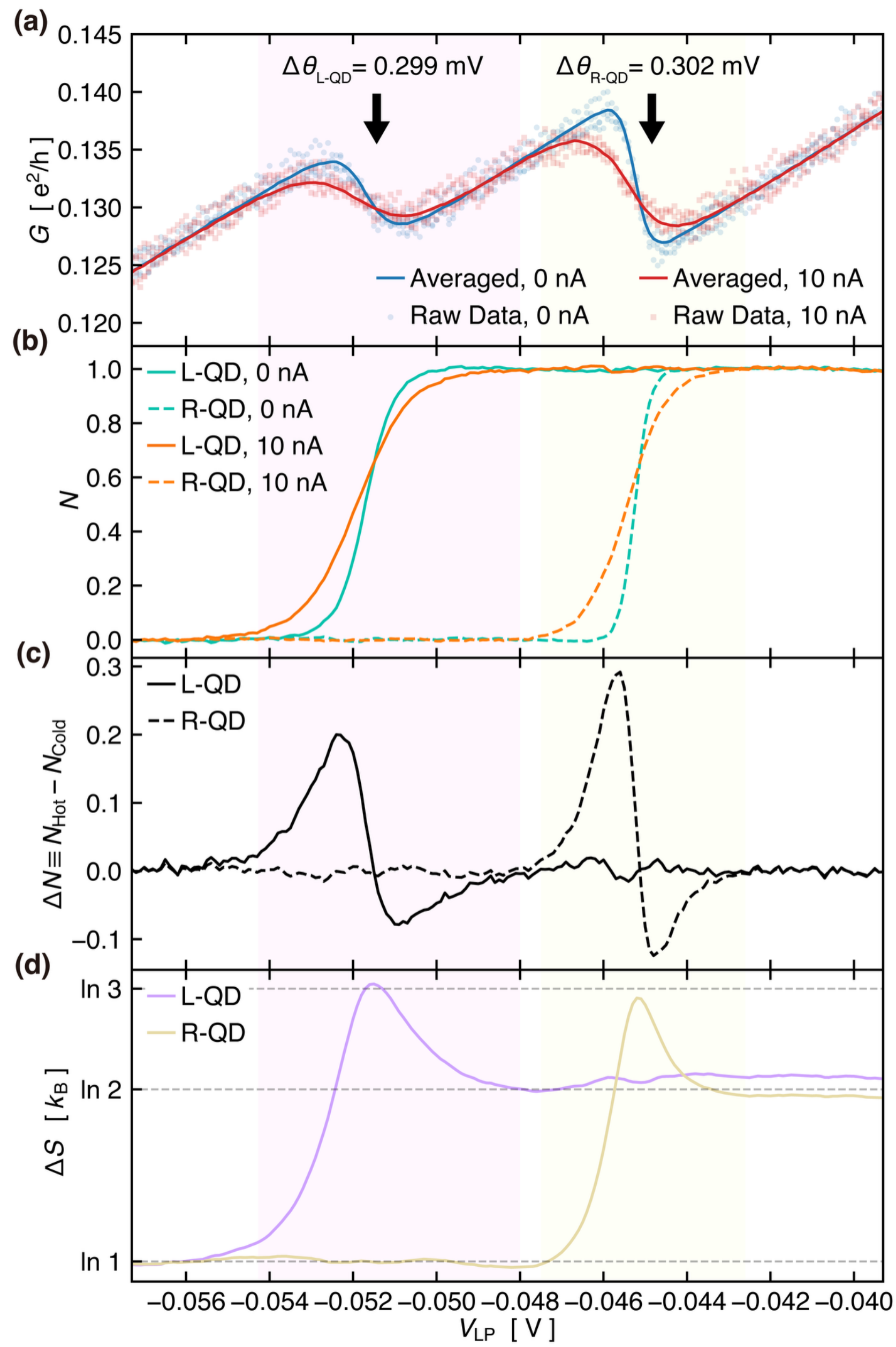

**Supplementary Figure S5. Entropy Extraction Process. (a)** Measured charge sensor (CS) conductance. This data set corresponds to Fig. 3(b–d) in the main text, with the measurement performed along the scan path indicated in Fig. 3(a). Blue and red points represent the raw data for the cold and hot conditions, respectively, while the corresponding solid lines denote the averaged signals. The change in thermal broadening, $\Delta\theta_i \equiv \theta_{i,Hot} - \theta_{i,\text{Cold}}$, for each transition is extracted by fitting the data to equation (S10). Here, the heating current for this measurement was 10 nA. **(b)** Occupation transition $N_i$ along the scan path. These traces are obtained by subtracting the fitted contributions of the adjacent transition and the linear background from the raw data. The solid and dashed lines represent the first and second transitions along the scan path, corresponding to the left (L-QD) and right (R-QD) quantum dots, respectively. **(c)** Change in occupation $\Delta N \equiv N_{\text{Hot}} - N_{\text{Cold}}$. Following the convention in (b), solid and dashed lines indicate the transitions of the L-QD and R-QD, respectively. **(d)** Calculated entropy from equation (S11).

The entropy extraction process is detailed in Fig. S5(a–d). This data set corresponds to Fig. 3(b–d) of the main text; the corresponding stability diagram and scan path are shown in Fig. 3(a). Initially, the CS signal is measured by sweeping both DQD plunger gates along the defined scan path. The same procedure is then repeated with a heating current applied. For the data shown in Fig. 3(d), a 10 nA current was supplied to both the left and right direct heating chambers, following the method described in the main text and Supplementary Note 4. The cold and hot signals are averaged separately, with the results displayed as blue and red solid lines in Fig. S5(a). Finally, each averaged signal is fitted to the following expression:

$$G_{\mathrm{CS}} = -\frac{g_1}{2}\tanh\left(\frac{V - V_{\mathrm{c1}}}{2\theta_1}\right) - \frac{g_2}{2}\tanh\left(\frac{V - V_{\mathrm{c2}}}{2\theta_2}\right) + \gamma_1 V + \gamma_2, \tag{S10}$$

where $g_i$, $\theta_i$, and $V_{\mathrm{c}i}$ ($i = 1,2$) represent the amplitude, thermal broadening (in voltage units), and center position of each charge transition, respectively, with $\gamma_1$ and $\gamma_2$ accounting for the linear background arising from capacitive coupling between the plunger gates and the CS. From these fits, we extract the change in thermal broadening $\Delta\theta_i = \theta_{i,\mathrm{Hot}} - \theta_{i,\mathrm{Cold}}$, which is later used in the entropy extraction process.

The next step is to obtain the occupation transition, $N_i(V)$, for each charge transition. This is achieved by subtracting the fitted contributions of the other transition and the linear background — specifically, $-\frac{g_j}{2}\tanh\left(\frac{V-V_{\mathrm{c}j}}{2\theta_j}\right) + \gamma_1 V + \gamma_2$, where $j \neq i$ — from the total signal, and then normalizing the result by $-g_i$. Figure S5(b) shows the obtained $N_i$. Here, since $i = 1$ and 2 correspond to the transitions of the L-QD and R-QD, respectively, they are labeled accordingly as the signals from the L-QD and R-QD. Then, $\Delta N_i \equiv N_{i,\mathrm{Hot}} - N_{i,\mathrm{Cold}}$ is calculated from $N_i$, as shown in Fig. S5(c).

Finally, $\Delta N_i$ is numerically integrated using a cumulative sum and normalized by the $\Delta\theta_i$ extracted previously (Fig. S5(a)). The resulting entropy change is given by:

$$\Delta S_i(V) = \frac{k_{\mathrm{B}}}{\Delta\theta_i}\sum_{v=V_0}^{V} \Delta N_i(v) \cdot \Delta V \tag{S11}$$

Here, $\Delta V$ is defined as the incremental step of a single gate voltage. This choice is made because the Maxwell relation—expressed as $\Delta S_{V_{i,1}\to V_{i,2}} = \int_{V_{i,1}}^{V_{i,2}} \alpha_{\delta,i}\frac{dN_{\mathrm{DQD}}}{dT_{\mathrm{e}}}dV_i$, where $\alpha_{\delta,i} \equiv -\partial\delta/\partial V_i$ —is formulated with respect to a specific plunger gate voltage. In our analysis, we specifically chose $V_{\mathrm{LP}}$ as the main gate for this integration process. After this numerical integration, the entropy evolution along the path is obtained as illustrated in Fig. S5(d).

## Supplementary Note 7: Single Quantum Dot Model in the Merged SQD Limit

In the merged SQD limit of Fig. 4(g–i), we describe the DQD system using an SQD model. The theoretical entropy shown in Fig. 4(i) is evaluated using the following SQD Hamiltonian[7,12]: $\hat{H}_{\mathrm{SQD}} = \sum_{\sigma=\uparrow,\downarrow} \varepsilon_{\mathrm{QD}}\, \hat{c}^{\dagger}_{\mathrm{QD},\sigma}\, \hat{c}_{\mathrm{QD},\sigma} + E_{\mathrm{C}}\, \hat{n}_{\mathrm{QD},\uparrow}\hat{n}_{\mathrm{QD},\downarrow}$. Here, $\varepsilon_{\mathrm{QD}}$ is the QD energy level, $E_{\mathrm{C}}$ is the charging energy of the merged SQD, the operator $\hat{c}^{\dagger}_{\mathrm{QD},\sigma}$ creates an electron with spin $\sigma$ in the energy level $\varepsilon_{\mathrm{QD}}$, and $\hat{n}_{\mathrm{QD},\sigma} = \hat{c}^{\dagger}_{\mathrm{QD},\sigma}\hat{c}_{\mathrm{QD},\sigma}$ is the corresponding number operator. Since only the empty, one-electron, and two-electron charge states are considered in the experiment, the merged SQD is modeled with a single level. This level is controlled by the plunger gate $V_{\mathrm{LP}}$ through the lever arm $\alpha$ along the $\delta$-axis as $\varepsilon_{\mathrm{QD}} = \frac{E_{\mathrm{C}}}{2} - \alpha(V_{\mathrm{LP}} - V_{\mathrm{L0}})$ where $V_{\mathrm{L0}}$ is the gate voltage offset that sets $\varepsilon_{\mathrm{QD}}$ relative to the reservoir chemical potential, which is taken as zero.

The lever arm $\alpha$ cannot be determined from bias spectroscopy because a bias triangle is no longer formed in this limit (Supplementary Note 1). We therefore express the SQD model in gate voltage units by dividing the Hamiltonian and the relevant energy scales by $\alpha$. Defining $\widetilde{H}_{\mathrm{SQD}} \equiv \frac{\hat{H}_{\mathrm{SQD}}}{\alpha}$, $\tilde{\varepsilon}_{\mathrm{QD}} \equiv \frac{\varepsilon_{\mathrm{QD}}}{\alpha}$, and $\tilde{E}_{\mathrm{C}} \equiv \frac{E_{\mathrm{C}}}{\alpha}$, the QD energy level in gate voltage units is written as $\tilde{\varepsilon}_{\mathrm{QD}} = \frac{\tilde{E}_{\mathrm{C}}}{2} - (V_{\mathrm{LP}} - V_{\mathrm{L0}})$.

The theoretical entropy is calculated using the Gibbs entropy $S(V_{\mathrm{LP}}; T_{\mathrm{e}}) = -k_{\mathrm{B}} \sum_i p_i \ln p_i$ where $p_i$ is the probability of finding the SQD in eigenstate $i$. We assume that the merged SQD remains in thermal equilibrium with the reservoirs and neglect hybridization between the SQD and the reservoirs because the QD–reservoir couplings are much smaller than the thermal energy scale. Under these assumptions, $p_i$ is given by $p_i(V_{\mathrm{LP}}; T_{\mathrm{e}}) = \frac{\exp\left(-\frac{\tilde{E}_i(V_{\mathrm{LP}})}{\theta}\right)}{Z(V_{\mathrm{LP}}; T_{\mathrm{e}})}$ where $\tilde{E}_i(V_{\mathrm{LP}})$ is the corresponding eigenenergy in gate voltage units, $Z(V_{\mathrm{LP}}; T_{\mathrm{e}}) = \sum_i \exp\left(-\frac{\tilde{E}_i(V_{LP})}{\theta}\right)$ is the partition function, and $\theta = \frac{k_{\mathrm{B}} T_{\mathrm{e}}}{\alpha}$ is the thermal energy scale in gate voltage units.

# Supplementary Note 8: Comparison Between 10 nA and 20 nA Heating Currents

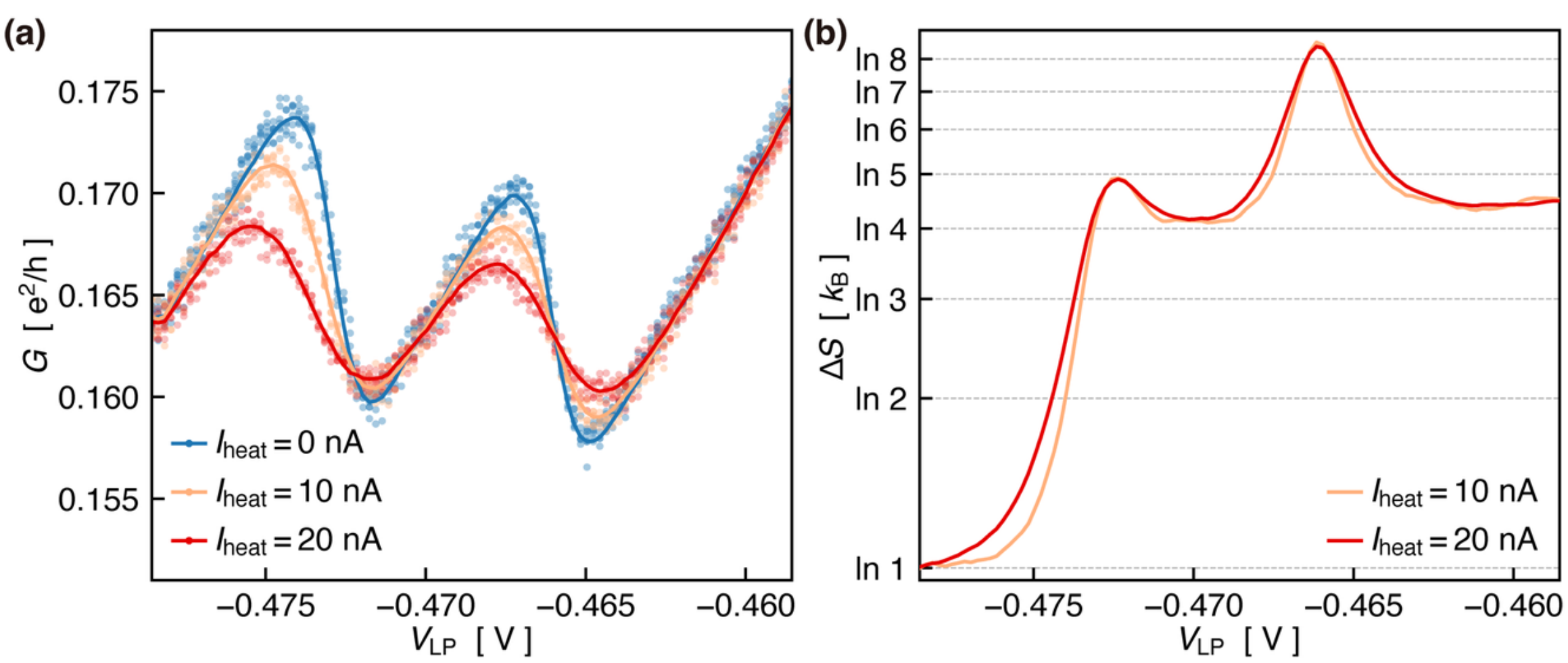

**Supplementary Figure S6. Heating Current Dependence of the Entropy. (a)** Raw data corresponding to Fig. 5(a) of the main text. **(b)** Entropy calculated using equation (S11) from the raw data presented in (a). For the $I_{\mathrm{heat}} = 10$ nA trace, the entropy is obtained by subtracting the 0 nA conductance trace from the 10 nA trace; a similar procedure is applied for the 20 nA case.

## Supplementary Note 9: Full Entropy Evolution Map

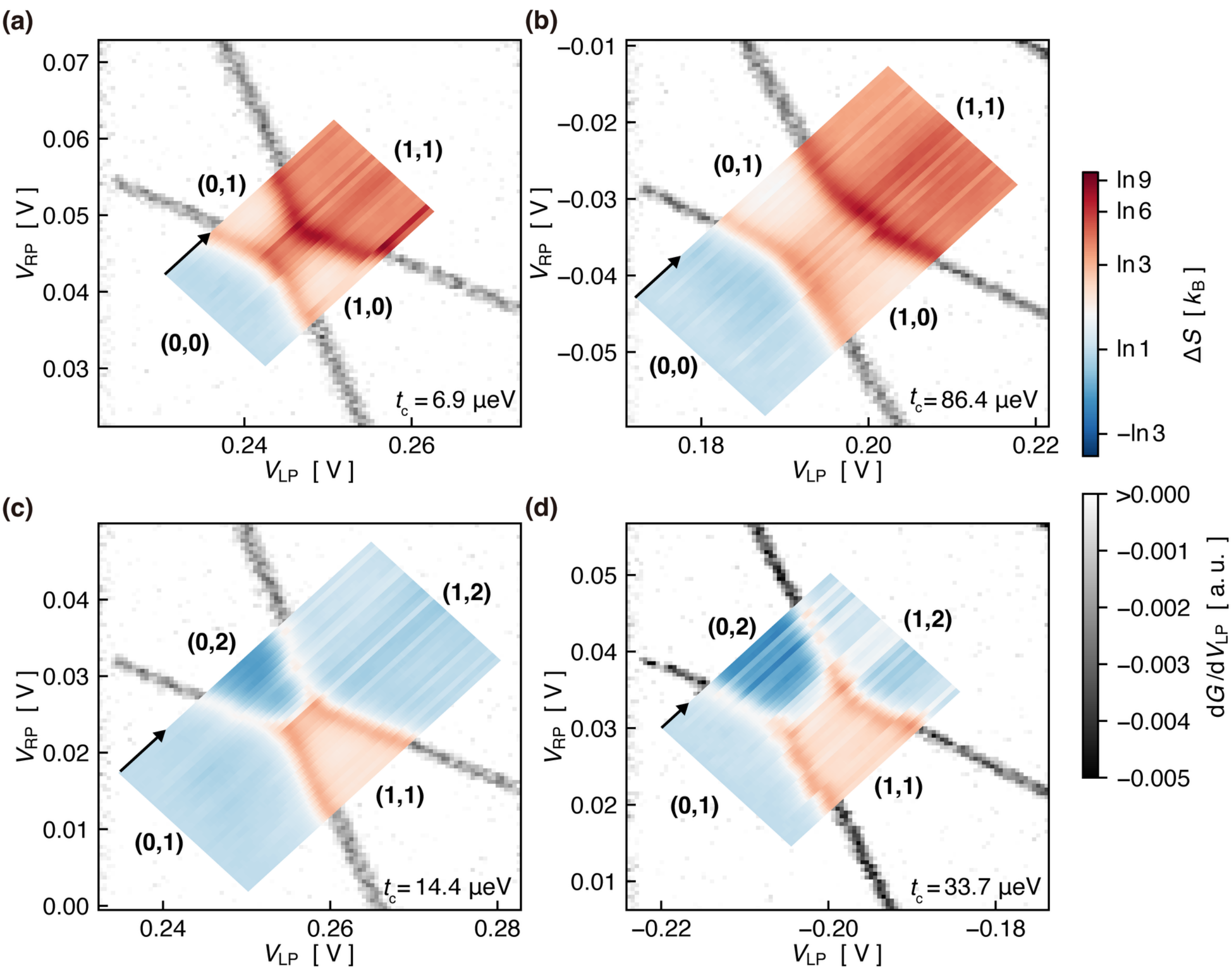

**Supplementary Figure S7. Full Entropy Evolution Map. (a, b)** Entropy evolution maps for the $(0,0) \rightarrow (1,1)$ transition with $t_c = 6.9$ μeV and $t_c = 86.4$ μeV, corresponding to the red and blue circles in Fig. 5(a) of the main text, respectively. **(c, d)** Entropy evolution maps for the $(0,1) \rightarrow (1,2)$ transition with $t_c = 14.4$ μeV and $t_c = 33.7$ μeV, corresponding to the red and blue squares in Fig. 5(c) of the main text, respectively. The grayscale colormap represents the charge sensor (CS) signal, with the entropy overlaid using a blue-to-red colormap. The arrows indicate the $\delta$-axis sweep direction.

## Supplementary Note 10: Width of Entropy along the $\varepsilon$-axis

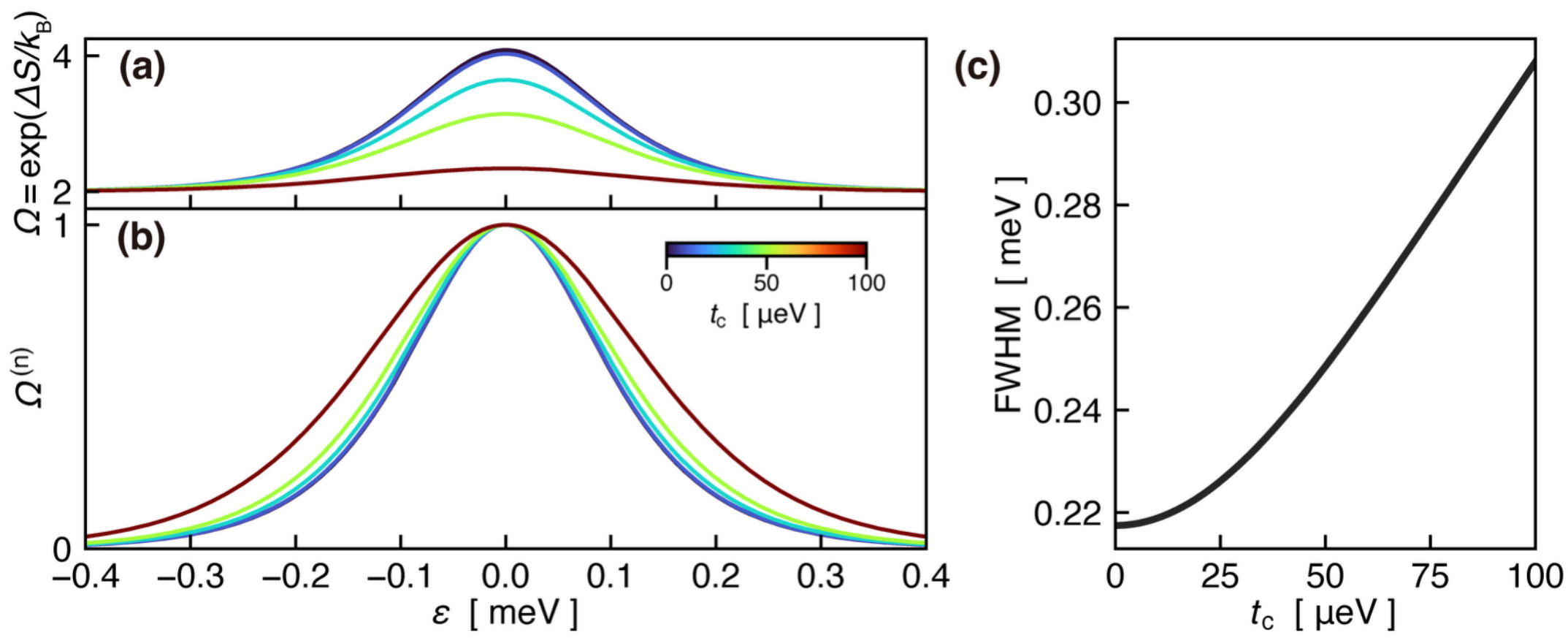

**Supplementary Figure S8. Width of Entropy along the $\varepsilon$-axis. (a)** Calculated number of thermally accessible states, $\Omega \equiv \exp[\Delta S(\varepsilon)/k_\mathrm{B}]$, with the other parameters fixed at the values used for the $t_\mathrm{c} = 86.4$ μeV case in main Fig. 5(a) (blue open circle). **(b)** Normalized $\Omega^{(n)}(\varepsilon)$. Panels (a) and (b) share the same colorbar. **(c)** Full width at half maximum (FWHM) extracted from the curves in (b).

Following the procedure described in the main text, we calculate the $\varepsilon$-dependent entropy $\Delta S(\varepsilon)$ using a series of $\delta$-axis line cuts for a broad range of $t_\mathrm{c}$, with all other parameters fixed at the values used for the $t_\mathrm{c} = 86.4$ μeV case in Fig. 5(a). Figure S8(a) shows the resulting number of thermally accessible states $\Omega \equiv \exp[\Delta S(\varepsilon)/k_\mathrm{B}]$. We normalize each curve according to $\Omega^{(n)}(\varepsilon) = \frac{\Omega(\varepsilon)-\Omega_{\min}}{\mathrm{Max}(\Omega(\varepsilon)-\Omega_{\min})}$ as shown in Fig. S8(b), and then extract the full width at half maximum (FWHM) of each normalized curve. As shown in Fig. S8(c), the FWHM increases monotonically with $t_\mathrm{c}$. Although the FWHM does not directly quantify the degree of hybridization between two localized charge states, it qualitatively reflects the detuning range over which the two QD levels remain close in energy to hybridize.

**References for Supplementary Information**